\documentclass[journal, final, onecolumn,12pt]{IEEEtran}  
\usepackage{cite}
\usepackage{graphicx}
\usepackage{subfigure}
\usepackage{mathrsfs}
\usepackage{amsmath}
\usepackage{amsfonts}
\usepackage{amssymb}

\newtheorem{theorem}{{Theorem}}
\newtheorem{lemma}[theorem]{{Lemma}}

\newtheorem{definition}{{Definition}}

\newcommand{\mb}{\mathbf}
\newcommand{\qed}{\hspace*{\fill} $\Box$ \\}

\begin{document}

\title{Wireless Network Resilience to Degree-Dependent and Cascading Node Failures}

\author{Zhenning~Kong,~\IEEEmembership{Student Member,~IEEE,}
        Edmund~M. Yeh,~\IEEEmembership{Member,~IEEE}\\
\thanks{This research is supported in part by National
Science Foundation (NSF) Cyber Trust grant CNS-0716335, and by Army Research Office (ARO) grant
W911NF-07-1-0524.}
\thanks{The material in this paper was presented in part at the Information Theory and Applications Workshop (ITA), San Diego, CA, Jan 2008.}
\thanks{Zhenning Kong and Edmund~M. Yeh are with the Department of Electrical Engineering, Yale University
(email: zhenning.kong@yale.edu, edmund.yeh@yale.edu)}}

\markboth{Submitted to \emph{IEEE Transactions on Information Theory}}{Submitted to \emph{IEEE
Transactions on Information Theory}}

\maketitle

\begin{abstract}
We study the problem of wireless network resilience to node failures from a percolation-based perspective. In practical
wireless networks, it is often the case that the failure probability of a node depends on its degree (number of
neighbors). We model this phenomenon as a degree-dependent site percolation process on random geometric graphs. Due to
its non-Poisson structure, degree-dependent site percolation is far from a trivial generalization of independent site
percolation. Using coupling and renormalization method, we obtain analytical conditions for the existence of phase
transitions within the degree-dependent failure model. Furthermore, in networks carrying traffic load, the failure of
one node can result in redistribution of the load onto other nearby nodes. If these nodes fail due to excessive load,
then this process can result in a cascading failure. Using a simple but descriptive model, we show that the cascading
failure problem for large-scale wireless networks is equivalent to a degree-dependent site percolation on random
geometric graphs. We obtain analytical conditions for cascades in this model. To our knowledge, this work represents
the first investigation of cascading phenomena in networks with geometric constraints.
\end{abstract}

\baselineskip 21 pt

\section{Introduction}

In large-scale wireless networks, nodes are often vulnerable to attacks, natural hazards, and resource depletion. The
ability of wireless networks to maintain global communication in the face of these challenges is a central concern for
network designers. For this purpose, a network may be considered to be functional if the size of the largest connected
component of operational nodes grows linearly with the size of the network. On the other hand, if the size of the
largest operational component vanishes as a fraction of the network as the network size grows, then the network is not
considered to be functional. A network may be said to be resilient if the remaining network is functional even after
many node and link failures. For instance, if the wireless sensor network still manages to collect information from a
constant fraction of the sensors even after a substantial number of node and link failures, then the network is
resilient. On the other hand, if after many node and link failures, the sensor network breaks down into isolated parts
where even the largest component can reach only a few other nodes, then the network is not considered to be resilient.
>From this perspective, the characterization of network resilience corresponds to the study of the qualitative and
quantitative properties of the largest connected component. A powerful tool for this study stems from the theory of
percolation~\cite{Gi61, MeRo96, Pe03, Gr99, BoRi06}. Recently, percolation theory, especially continuum percolation,
has been widely used to study the coverage, connectivity, and capacity of large-scale wireless networks~\cite{GuKu98,
BoNrFrMe03, FrBoCoBrMe05, DoMaTh04, DoFrTh05, DoBaTh05, DoFrMaMeTh06, KoYe07-4, KoYe08-1, KoYe08-2}.

A percolation process resides in a random graph structure, where nodes or links are randomly designated as either
``occupied'' or ``unoccupied.'' When the graph structure resides in continuous space, the resulting model is described
by continuum percolation~\cite{Gi61, Pe03, MeRo96}. A major focus of continuum percolation theory is the random
geometric graph induced by a Poisson point process with constant density $\lambda$. A fundamental result for continuum
percolation concerns a phase transition effect whereby the macroscopic behavior of the system is very different for
densities below and above some critical value $\lambda_c$. For $\lambda<\lambda_c$ (subcritical or non-percolated), the
connected component containing the origin (or any other fixed point) contains a finite number of points almost surely.
For $\lambda>\lambda_c$ (supercritical or percolated), the connected component containing the origin (or any other
fixed point) contains an infinite number of points with a positive probability~\cite{Gi61, MeRo96, Pe03, Gr99}.

In this paper, we study the resilience of large-scale wireless networks to node failures from the
percolation perspective. We first consider wireless networks with random, independent node
failures.  To see why this problem can be described by a percolation process on the network, note
that in a network with random node failures, nodes are randomly occupied (operational) or
unoccupied (failed), and the number of operational nodes that can successfully communicate with an
extensive portion of the network is precisely the largest component of the corresponding
percolation model. Hence, the phase transition phenomena of the percolation model directly
translates to a description of the random failures model.

In practical wireless networks, it is often the case that the failure probability of a node
depends on its degree (number of neighbors). For instance, a wireless sensor node which must
communicate with a large number of neighbors is more likely to deplete its energy reserve. A
communication node directly connected to many other nodes in a military network is more likely to
be attacked by an enemy seeking to break down the whole network.  Such phenomenon can be described
by a general model where each node fails with a probability depending on its degree.  In this
paper, we study such \emph{degree-dependent node failure} problems.  Specifically, by analyzing
the problem as a degree-dependent site percolation process on random geometric graphs, we obtain
analytical conditions on percolation in this model.

In networks which carry load, distribute a resource or aggregate data, such as wireless sensor networks and electrical
power networks, the failure of one node often results in redistribution of the load from the failed node to other
nearby nodes. If nodes fail when the load on them exceeds some maximum capacity or when the battery energy is depleted,
then a cascading failure or avalanche may occur because the redistribution of the load causes other nodes to exceed
their thresholds and fail, thereby leading to a further redistribution of the load. An example of such a cascading
failure is the power outage in the western United States in August 1996, which resulted from the spread of a small
initial power shutdown in El Paso, Texas. The power outage spread through six states as far as Oregon and California,
leaving several million customers without electronic power~\cite{KoTaMi99, SaCaLy00}. Cascades have also been studied
in social networks~\cite{Wa02, Ne03}. In wireless sensor networks constrained by battery resource, the system may
suffer similar cascading failure problems, though the cascading process may be much slower than that for power
networks. In this paper, we study cascade failures in large-scale wireless networks. To our knowledge, this is the
first work to address cascading phenomena in networks with geometric constraints. We show that such problems can be
mapped to a percolation process on random geometric graphs. Using our degree-dependent site percolation model, we
obtain analytical conditions on the occurrence of a cascading failure.

This paper is organized as follows. In Section II, we outline some preliminary results for random
geometric graphs and continuum percolation. In Section III, we first review independent random
node failures, and then study the general degree-dependent node failures problem. We provide
analytical conditions for the existence of an infinite component in these models. In Section IV,
we show the equivalence between cascading failure in large-scale wireless networks and
degree-dependent percolation, and investigate the conditions under which a small exogenous event
can trigger a global cascading failure. In Section V, we present simulation results, and finally,
we conclude in Section VI.

\section{Random Geometric Graphs and Continuum Percolation}

We use random geometric graphs to model wireless networks. That is, we assume that the network
nodes are randomly placed over some area or volume, and a communication link exists between two
(randomly placed) nodes if the distance between them is sufficiently small, so that the received
power is large enough for successful decoding. A mathematical model for this is as follows. Let
$\|\cdot\|$ be the Euclidean norm, and $f(\cdot)$ be some probability density function (p.d.f.) on
$\mathbb{R}^d$. Let ${\mb X}_1, {\mb X}_2, ..., {\mb X}_n$ be independent and identically
distributed (i.i.d.) $d$-dimensional random variables with common density $f(\cdot)$, where ${\mb
X}_i$ denotes the random location of node $i$ in $\mathbb{R}^d$. The ensemble of graphs with
undirected links connecting all those pairs $\{{\mb x}_i, {\mb x}_j\}$ with $\|{\mb x}_i- {\mb
x}_j \|\leq r, r>0,$ is called a \emph{random geometric graph}~\cite{Pe03}, denoted by $G({\cal
X}_n, r)$. The parameter $r$ is called the characteristic radius.

In the following, we consider random geometric graphs $G({\cal X}_n, r)$ in $\mathbb{R}^2$, with ${\mb X}_1, {\mb X}_2,
..., {\mb X}_n$ distributed i.i.d. according to a uniform distribution in the square ${\cal
A}=[0,\sqrt{\frac{n}{\lambda}}]^2$. Let $A=|{\cal A}|$ be the area of ${\cal A}$. In this case, ignoring border
effects, as $n\rightarrow \infty$ and $A \rightarrow \infty$ with $\frac{n}{A}=\lambda$ fixed, $G({\cal X}_n, r)$
converges to an infinite random geometric graph $G(\mathcal{H}_{\lambda},r)$ induced by a homogeneous Poisson point
process with density $\lambda>0$.\footnote{More precisely, this convergence is in distribution since Binomial
distribution converges to Poisson distribution.} Due to the scaling property of random geometric
graphs~\cite{MeRo96,Pe03}, in the following, we focus on $G(\mathcal{H}_{\lambda},1)$.

Consider a graph $G=(V,E)$, where $V$ and $E$ denote the set of nodes and links, respectively.
Given $u,v\in V$, we say $u$ and $v$ are \emph{adjacent} if there exists a link between $u$ and
$v$, i.e., $(u,v)\in E$. In this case, we also say that $u$ and $v$ are \emph{neighbors}.

Let $\mathcal{H}_{\lambda,\mathbf{0}}=\mathcal{H}_{\lambda}\cup \{\mathbf{0}\}$, i.e., the union
of the origin and the infinite homogeneous Poisson point process with density $\lambda$. Note that
in a random geometric graph induced by a homogeneous Poisson point process, the choice of the
origin can be arbitrary. As discussed before, a phase transition takes place at the critical
density. More formally, we have the following definition:

\vspace{0.1in}%
\begin{definition} For $G(\mathcal{H}_{\lambda,\mathbf{0}},1)$, the
\emph{percolation probability} $p_{\infty}(\lambda)$ is the probability that the component
containing the origin has an infinite number of nodes of the graph. The \emph{critical density}
$\lambda_c$ is defined as
\begin{equation}
\lambda_c=\inf \{\lambda>0: p_{\infty}(\lambda)>0\}.
\end{equation}
\end{definition}
\vspace{0.1in}%

It is known that if $\lambda>\lambda_c$, then there exists a unique infinite component in $G(\mathcal{H}_{\lambda},1)$.
A fundamental result of continuum percolation states that $0<\lambda_c<\infty$~\cite{MeRo96}. Exact values of
$\lambda_c$ and $p_{\infty}(\lambda)$ are not yet known. Simulation studies show that~$1.43<\lambda_c<1.44$
\cite{QuToZi00}.

\section{Random Node Failures}

\subsection{Independent Random Node Failures}

As we mentioned in the introduction, the problem of network resilience to random node failures can be described by a
percolation process on the graph modelling the network. Suppose the network modelled by $G(\mathcal{H}_{\lambda},1)$ is
subject to random node failures where each node fails, along with all associated links, with probability $q$,
independently of other nodes. When $q$ stays below a certain threshold $q_c$, there still exists a connected component
of operational nodes that spans the entire network. When $q > q_c$, the network disintegrates into smaller,
disconnected operational parts. Since each node fails randomly and independently with probability $q$, according to
Thinning Theorem~\cite{MeRo96, Pe03}, the remaining graph is still a random geometric graph with density
$(1-q)\lambda$. Thus, given $\lambda>\lambda_c$, the remaining graph is percolated if $(1-q)\lambda>\lambda_c$, and not
percolated if $(1-q)\lambda<\lambda_c$. Therefore, we have
\begin{equation}\label{eta-c}
q_c = 1 -\frac{\mu_c}{\mu} = 1-\frac{\lambda_c}{\lambda},
\end{equation}
where $\mu_c$ ($\mu_c=\lambda_c\pi$) and $\mu$ are the critical mean degree and the mean degree of
$G(\mathcal{H}_{\lambda},1)$, respectively.

\subsection{Degree-Dependent Node Failures}

We have thus far considered wireless network resilience to independent random node failures. As we
mentioned before, in practical wireless networks, it is often the case that the failure
probability of a node depends on its degree. We therefore study network resilience in the face of
degree-dependent node failures.  Let the original random geometric graph be
$G(\mathcal{H}_\lambda,1)$ with density $\lambda>\lambda_c$. Suppose each node with degree $k$ in
$G(\mathcal{H}_\lambda,1)$ fails, along with all associated links, with probability $q(k), 0\leq
q(k) \leq 1$. Denote the remaining graph consisting of operational nodes and associated links by
$G(\mathcal{H}_\lambda,1,q(\cdot))$. We say $G(\mathcal{H}_\lambda,1,q(\cdot))$ is percolated if
there exists an infinite component in $G(\mathcal{H}_\lambda,1,q(\cdot))$.

Note that in wireless networks, a node with more neighbors (higher degree $k$) may suffer from more interference. If we
take the failure probability $q(k)$ to be increasing in $k$, then the effects of interference can be captured by our
failure model.

To study the percolation-based connectivity of $G(\mathcal{H}_\lambda,1,q(\cdot))$, we consider a
degree-dependent site percolation process for random geometric graphs. Similar problems have been
studied in the context of Erd\"{o}s-Renyi random graphs and random graphs with given degree
distributions using generating function methods~\cite{Ne03, CoErAvHa00, CaNeStWa00, NeStWa01}. Due
to clustering effects and geometric constraints, however, generating function methods are not
applicable for random geometric graphs. The SINR-based percolation model for wireless networks
studied in~\cite{DoBaTh05, DoFrMaMeTh06} involve dependent percolation but not degree-dependent
percolation. In~\cite{GoRa04}, a degree-dependent site percolation model is studied. There, the
authors propose a topology control mechanism for sensor networks where each sensor stays active
for a $\frac{\phi}{k}$ fraction of the time, where $\phi$ is a constant and $k>\phi$. The authors
obtain a sufficient condition for the existence of an infinite component within this model. A more
general model is studied in~\cite{KoYe07-4}. As in~\cite{GoRa04}, the authors in~\cite{KoYe07-4}
obtain only a sufficient condition for the existence of an infinite component. In this paper, in
addition to a sufficient condition, a necessary condition for the existence of an infinite
component is found for our model. The main results are as follows.

\vspace{0.1in}%
\begin{theorem}\label{Theorem-General-Degree-Dependent}
(i) For any $\mu_1>\mu_c$ and $G(\mathcal{H}_\lambda,1)$ with $\mu>\mu_1$, there exists $k_0<
\infty$ which depends on $\mu$, such that if
\begin{equation}\label{qk-upper-bound}
q(k)\leq 1-\frac{\mu_1}{\mu}, \quad \mbox{for all}~1\leq k \leq k_0,
\end{equation}
then with probability 1, there exists an infinite connected component in
$G(\mathcal{H}_\lambda,1,q(\cdot))$;

(ii) Given $G(\mathcal{H}_\lambda,1)$ with $\lambda>\lambda_c$, if either
\begin{equation}\label{qk-lower-bound-1} e^{-\frac{\lambda}{2}}+\sum_{k=1}^{\infty}
\frac{(\frac{\lambda}{2})^k}{k!}e^{-\frac{\lambda}{2}}q(k-1)^k>1-\frac{1}{27}
\end{equation}
when $q(k)$ is non-decreasing in $k$, or if
\begin{equation}\label{qk-lower-bound-2}
\sum_{k=1}^{\infty}\frac{\left(\frac{\lambda}{2}\right)^k}{k!}e^{-\frac{\lambda}{2}}
\sum_{m=0}^{\infty}\frac{[\lambda(2\sqrt{2}+\pi)]^m}{m!}e^{-\lambda(2\sqrt{2}+\pi)}
\left(1-q(m+k-1)^k\right)<\frac{1}{27}
\end{equation}
when $q(k)$ is non-increasing in $k$, then with probability 1, there is no infinite connected
component in $G(\mathcal{H}_\lambda,1,q(\cdot))$.
\end{theorem}
\vspace{0.1in}%

An interesting implication of Theorem \ref{Theorem-General-Degree-Dependent}-(i) is that even if
all nodes with degree larger than $k_0$ fail with probability 1, an infinite component still
exists in the remaining graph as long as~\eqref{qk-upper-bound} is satisfied.

Note that although~\eqref{qk-upper-bound} resembles the percolation condition for independent node failures, Theorem
\ref{Theorem-General-Degree-Dependent} is far from a straightforward generalization of the result for independent
failures. Indeed, in the degree-dependent model, for general $q(k)$, the spatial distribution of the operational nodes
(or failed nodes) is no longer homogeneous Poisson or even nonhomogeneous Poisson. Nevertheless, if the resulting point
process dominates the Poisson point process with critical density in the sense that
\[
\int_{\mathcal{A}}\lambda(\mathbf{x})d\mathbf{x} > \lambda_c|\mathcal{A}|
\]
for every area $\mathcal{A}\subset\mathbb{R}^2$, where $\lambda_c$ is the critical density of the Poisson point process
and $\lambda(\mathbf{x})$ is the density function of the point process resulting from the degree-dependent failure
model,\footnote{Precisely, given the point process resulting from the degree-dependent failure model,
$\lambda(\mathbf{x})=\lim_{\delta\rightarrow 0}\Pr(\exists \mbox{ one node} \in \mathcal{A}(\mathbf{x},\delta))$, where
$\mathcal{A}(\mathbf{x},\delta)$ is the circular region centered at $\mathbf{x}$ with radius $\delta$.} then using
Strassen's Theorem~\cite{St65}, we can couple the two point processes to show that the resulting graph is always
percolated. Given the general form of $q(k)$, however, computing the density function of the resulting point process is
difficult.

To tackle this problem, we use a renormalization argument that employs a mapping between the continuum model and a
discrete percolation model. A similar technique was used in \cite{DoFrTh05, DoFrMaMeTh06}. Using the fact that this
mapping is one to one, we can bound the density of the point process resulting from the degree-dependent failure model,
and then resort to coupling methods. In particular, we will couple $G(\mathcal{H}_\lambda, 1, q(\cdot))$ with another
random failure model which is percolated. We will show that when~\eqref{qk-upper-bound} is satisfied, there exists
$k_0<\infty$ such that all the operational nodes having degree less than or equal to $k_0$ in the random failure model
are operational in $G(\mathcal{H}_\lambda, 1, q(\cdot))$, and these operational nodes form an infinite component in
$G(\mathcal{H}_\lambda, 1, q(\cdot))$.

\vspace{+0.1in}%
\emph{Proof of Theorem \ref{Theorem-General-Degree-Dependent}-(i):} To prove Theorem
\ref{Theorem-General-Degree-Dependent}-(i), consider a square lattice
$\mathcal{L}=d\cdot\mathbb{Z}^2$, where $d$ is the edge length. The vertices of $\mathcal{L}$ are
located at $(d\times i, d\times j)$ where $(i,j)\in \mathbb{Z}^2$. For each horizontal edge $a$,
let the two end vertices be $(d\times a_x, d\times a_y)$ and $(d\times a_x+d, d\times a_y)$.

Now consider a random failure model in $G(\mathcal{H}_{\lambda},1)$ where each node fails (with
all associated links) independently with probability $1-\frac{\mu_1}{\mu}$. Let
$G_1(\mathcal{H}_\lambda, 1)$ be the remaining graph. By the Thinning Theorem,
$G_1(\mathcal{H}_\lambda, 1)$ is a random geometric graph with density
$\lambda_1=\frac{\mu_1}{\pi}>\lambda_c$. Consequently, $G_1(\mathcal{H}_\lambda, 1)$ is in the
supercritical regime.

\begin{figure}[t!]
\centerline{ \subfigure[Good Rectangle]{
\includegraphics[width=3in]{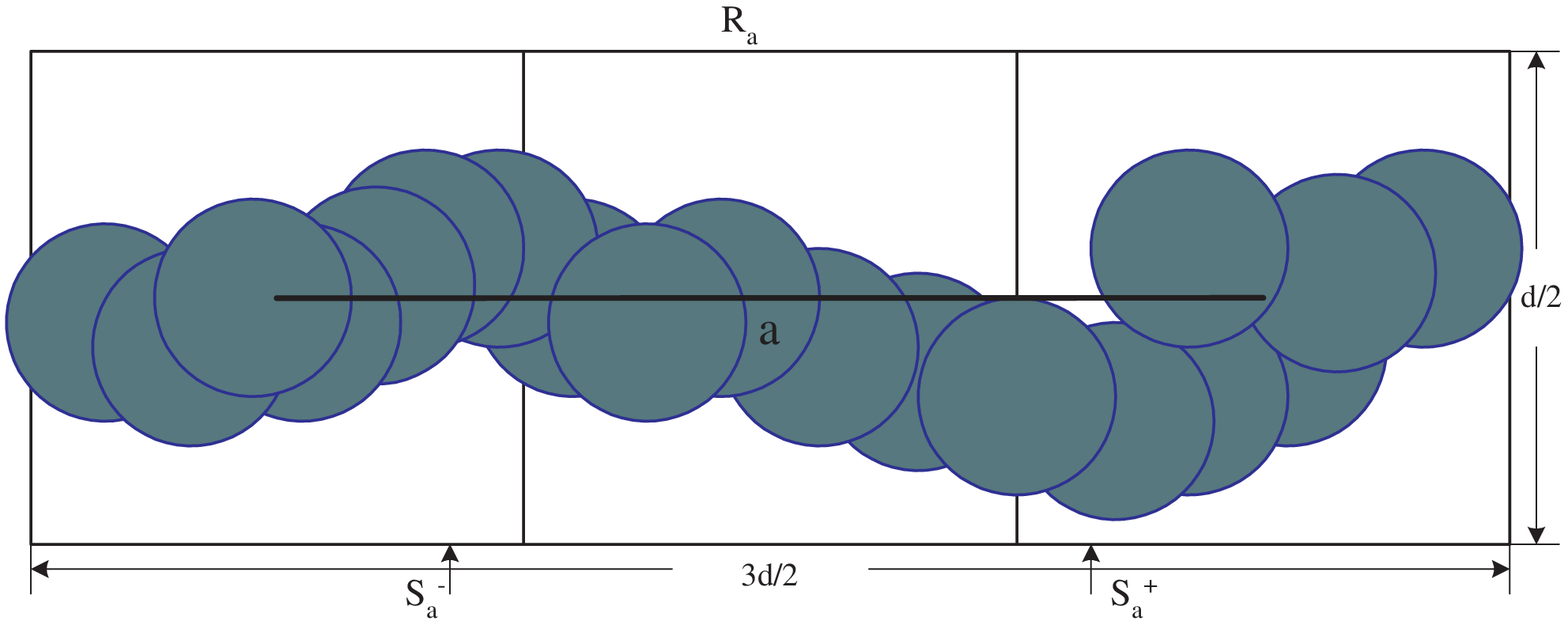}}\hfil
\subfigure[Complete Rectangle]{
\includegraphics[width=3in]{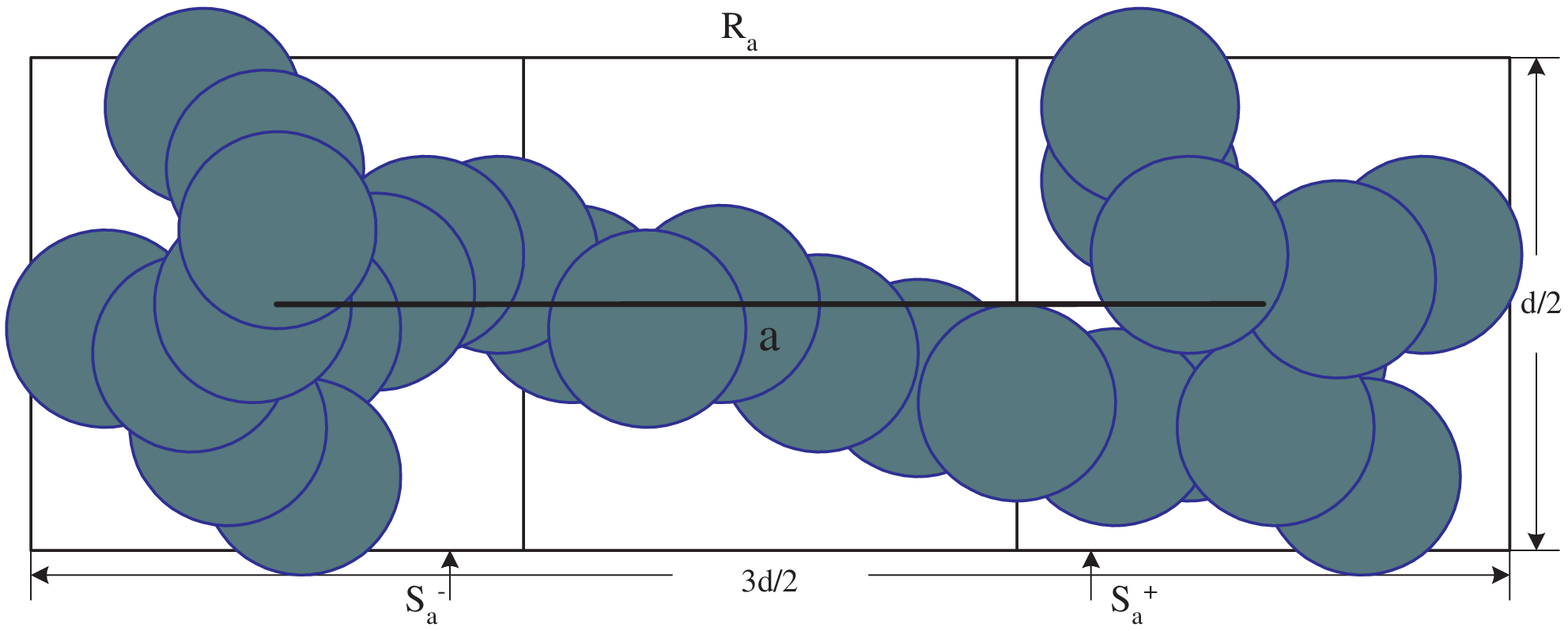}}}
\caption{Examples of good and complete rectangles (edges)}\label{fig:GoodOpenRectangle}
\end{figure}

Define event $A_a(d)$ for edge $a$ in $\mathcal{L}$ as the set of outcomes for which the following
condition is satisfied: the rectangle $R_a=[a_xd-\frac{d}{4},a_xd+\frac{5d}{4}]\times
[a_yd-\frac{d}{4},a_yd+\frac{d}{4}]$ is crossed\footnote{Here, a rectangle
$R=[x_1,x_2]\times[y_1,y_2]$ being crossed from left to right by a connected component in
$G_1(\mathcal{H}_\lambda,1)$ means that there exists a sequence of nodes $v_1,v_2,...,v_m\in
G_1(\mathcal{H}_\lambda,1)$ contained in $R$, with $||\mathbf{x}_{v_i}-\mathbf{x}_{v_{i+1}}||\leq
1, i=1,...,m-1$, and $0<x(v_1)-x_1<1, 0<x_2-x(v_m)<1$, where $x(v_1)$ and $x(v_m)$ are the
$x$-coordinates of nodes $v_1$ and $v_m$, respectively.  A rectangle being crossed from top to
bottom is defined analogously.} from left to right by a connected component in
$G_1(\mathcal{H}_\lambda,1)$. If $A_a(d)$ occurs, we say that rectangle $R_a$ is a {\em good}
rectangle, and edge $a$ is a {\em good} edge. Let
\[
p_g(d)\triangleq\Pr(A_a(d)).
\]
Define $A_a(d)$ similarly for all vertical edges by rotating the rectangle by $90^{\circ}$. An
example of a good rectangle and a good edge is illustrated in
Figure~\ref{fig:GoodOpenRectangle}-(a).

Further define event $B_a(d)$ for edge $a$ in $\mathcal{L}$ as the set of outcomes for which both
of the following occur:
\begin{itemize}
\item[(i)] $A_a(d)$ occurs; \item[(ii)] The left square
$S_a^-=[a_xd-\frac{d}{4},a_xd+\frac{d}{4}]\times [a_yd-\frac{d}{4},a_yd+\frac{d}{4}]$ and the
right square $S_a^+=[a_xd+\frac{3d}{4},a_xd+\frac{5d}{4}]\times
[a_yd-\frac{d}{4},a_yd+\frac{d}{4}]$ are both crossed from top to bottom by connected components
in $G_1(\mathcal{H}_\lambda,1)$.
\end{itemize}
If $B_a(d)$ occurs, we say that rectangle $R_a$ is a {\em complete} rectangle, and edge $a$ is a
{\em complete} edge. Let
\[
p_c(d)\triangleq\Pr(B_a(d)).
\]
Define $B_a(d)$ similarly for all vertical edges by rotating the rectangle by $90^{\circ}$. An
example of a complete rectangle and a complete edge is illustrated in
Figure~\ref{fig:GoodOpenRectangle}-(b).

Note that the events $\{B_a(d)\}$ are not independent in general. However, if two edges $a$ and
$b$ are not adjacent, i.e., they do not share any common end vertices, then $B_a(d)$ and $B_b(d)$
are independent.

\begin{figure}[t]
\centering
\includegraphics[width=2in]{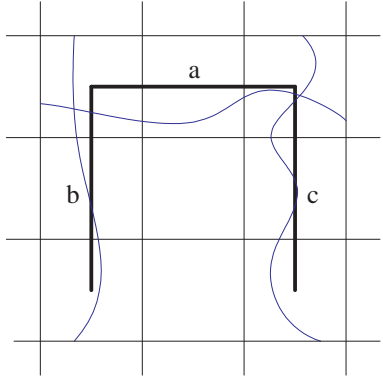}
\caption{Events $A_a(d)$, $A_b(d)$ and $A_c(d)$ imply event $B_a(d)$.}\label{fig:GoodOpen}
\end{figure}

As illustrated in Figure~\ref{fig:GoodOpen}, edges $b$ and $c$ are vertically adjacent to edge $a$. It is clear that
when events $A_a(d)$, $A_b(d)$ and $A_c(d)$ occur, event $B_a(d)$ occurs. Moreover, since events $A_a(d)$, $A_b(d)$ and
$A_c(d)$ are increasing events\footnote{An event $A$ is called increasing if $I_A(G)\leq I_A(G')$ whenever graph $G$ is
a subgraph of $G'$, where $I_A$ is the indicator function of $A$. An event $A$ is called decreasing if $A^{c}$ is
increasing. For details, please see~\cite{Gr99, MeRo96, Pe03}.}, according to the Fortuin-Kasteleyn-Ginibre (FKG)
inequality~\cite{Gr99, MeRo96, Pe03},
\begin{eqnarray*}
p_c(d) & = & \Pr(B_a(d))\\
&\geq&\Pr(A_a(d)\cap A_b(d)\cap A_c(d))\\
&\geq&\Pr(A_a(d))\Pr(A_b(d))\Pr(A_c(d))\\
& = &(p_g(d))^3.
\end{eqnarray*}

According to Corollary 4.1 in~\cite{MeRo96}, the probability $p_g(d)$ converges to 1 as
$d\rightarrow \infty$ when $G_1(\mathcal{H}_{\lambda},1)$ is in the supercritical phase. In this
case, $(p_g(d))^3$ converges to 1 as $d\rightarrow \infty$ as well. Hence, $p_c(d)$ converges to 1
as $d\rightarrow \infty$ when $G_1(\mathcal{H}_{\lambda},1)$ is in the supercritical phase.

\textbf{Now, define}
\begin{equation}\label{d-epsilon}
d(\lambda)\triangleq
\inf\left\{d>4:p_c(d)-\frac{1}{\left(\frac{d}{2}+2\right)\left(\frac{3d}{2}+2\right)
\lambda}>1-q_0\right\},
\end{equation}
where $q_0\triangleq \frac{1}{9+2\sqrt{3}}$. Now choose the edge length of $\mathcal{L}$ as
$d=d(\lambda)$. We further define complete events $\{B_a'(d)\}$ with respect to
$G(\mathcal{H}_\lambda, 1, q(\cdot))$ in the same way as we defined complete events $\{B_a(d)\}$
with respect to $G_1(\mathcal{H}_\lambda, 1)$.

\begin{figure}[t]
\centering
\includegraphics[width=4in]{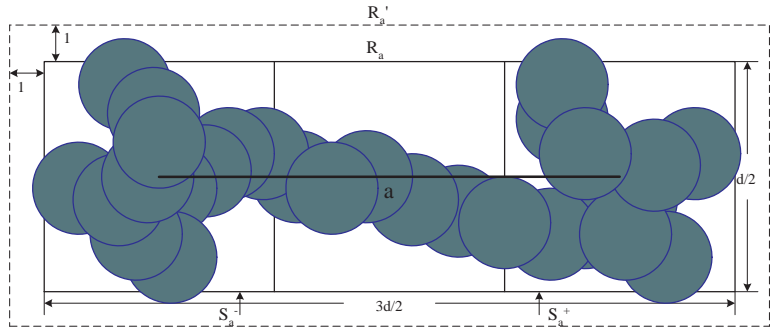}
\caption{Rectangle $R_a'$ is the rectangle $R_a$ extended by 1 in all
directions.}\label{fig:GoodRectangleNew}
\end{figure}

Define event $C_a(d)$ for each horizontal edge $a$ in $\mathcal{L}$ as the set of outcomes for
which the following condition is satisfied: The number of nodes of $G(\mathcal{H}_\lambda, 1)$ in
$R_a'$ is strictly less than
\begin{equation}\label{N-1}
k_0\triangleq 2\left(\frac{d(\lambda)}{2}+2\right)\left(\frac{3d(\lambda)}{2}+2\right)\lambda,
\end{equation}
where $R_a'=[a_xd(\lambda)-\frac{d(\lambda)}{4}-1,a_xd+\frac{5d(\lambda)}{4}+1]\times
[a_yd(\lambda)-\frac{d(\lambda)}{4}-1,a_yd(\lambda)+\frac{d(\lambda)}{4}+1]$, i.e., $R_a'$ is the
rectangle $R_a$ extended by 1 in all directions. Rectangle $R_a'$ is shown in
Figure~\ref{fig:GoodRectangleNew}. Note that
$|R_a'|=\left(\frac{d(\lambda)}{2}+2\right)\left(\frac{3d(\lambda)}{2}+2\right)$.

Define $C_a(d)$ similarly for all vertical edges by rotating the rectangle by $90^{\circ}$. If
$C_a(d)$ occurs, we call rectangle $R_a$ and edge $a$ {\em efficient}. Let
\[
p_e(d)\triangleq\Pr(C_a(d)).
\]
Note that the events $\{C_a(d)\}$ are not independent in general due to potential overlaps. However, if $d>4$ and two edges
$a$ and $b$ are not adjacent, i.e., they do not share any common end vertices, then $C_a(d)$ and
$C_b(d)$ are independent.

We say an edge $a$ in $\mathcal{L}$ is {\em open} if and only if it is both complete and
efficient, i.e., when events $B_a(d)$ and $C_a(d)$ both occur, and {\em closed} otherwise.

\begin{figure}[t!]
\centering
\includegraphics[width=2.5in]{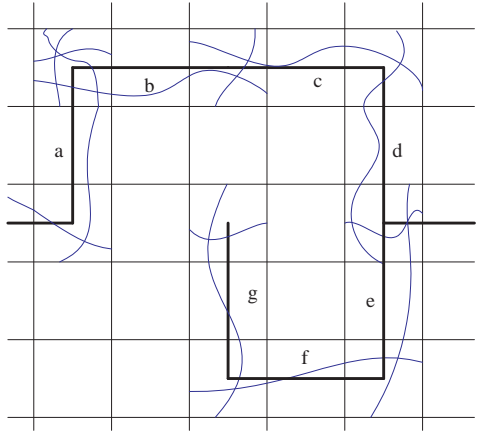}
\caption{A path of open edges in $\mathcal{L}$ implies a path of connected component in
$G(\mathcal{H}_\lambda,1,q(\cdot))$}\label{fig:GoodPathNew}
\end{figure}

When $C_a(d)$ occurs for edge $a$ in $\mathcal{L}$, no node of $G(\mathcal{H}_\lambda,1,q(\cdot))$
in $R_a$ has degree strictly greater than $k_0$ in $G(\mathcal{H}_\lambda,1)$. In addition, if
$q(k)$ satisfies (\ref{qk-upper-bound}), a node in $G(\mathcal{H}_\lambda,1)$ with degree $k$,
$1\leq k\leq k_0$, survives with a probability greater than or equal to $\frac{\mu_1}{\mu}$ in the
degree-dependent failures model. On the other hand, for the independent random failures model, a
node in $G(\mathcal{H}_\lambda,1)$ survives with probability exactly equal to $\frac{\mu_1}{\mu}$.
Thus we can couple $G(\mathcal{H}_\lambda,1,q(\cdot))$ with $G_1(\mathcal{H}_\lambda, 1)$ so that
the existence of crossings defined in events $\{B_a(d)\}$ for $G_1(\mathcal{H}_\lambda, 1)$
implies the existence of crossings defined in events $\{B_a'(d)\}$ for
$G(\mathcal{H}_\lambda,1,q(\cdot))$. Hence, if edge $a$ of $\mathcal{L}$ is open, there exists at
least one left-to-right crossing and two top-to-bottom crossings in $R_a$ in
$G(\mathcal{H}_\lambda,1,q(\cdot))$. Therefore, a path of open edges in $\mathcal{L}$ implies a
connected component in $G(\mathcal{H}_\lambda,1,q(\cdot))$. This is illustrated in
Figure~\ref{fig:GoodPathNew}.

Although events $B_a(d)$ and $C_a(d)$ are not independent, we have
\begin{eqnarray}\label{p-open}
p_o(d) & \triangleq &\Pr(B_a(d)\cap C_a(d))\nonumber\\
& = & \Pr(B_a(d))+\Pr(C_a(d))-\Pr(B_a(d)\cup C_a(d))\nonumber \\
& \geq & p_c(d)+p_e(d)-1.
\end{eqnarray}

Let $N$ be the number of nodes of $G(\mathcal{H}_\lambda, 1)$ in $R_a'$. Then $N$ has a Poisson
distribution with mean
\[
E[N]=\left(\frac{d(\lambda)}{2}+2\right) \left(\frac{3d(\lambda)}{2}+2\right)\lambda.
\]
Note that $k_0=2E[N]$. By Chebychev's inequality, we have
\begin{eqnarray}\label{p-efficient-bound}
p_e(d) & = & \Pr(N< k_0)\nonumber\\
& = & 1 - \Pr(N\geq k_0)\nonumber\\
& = & 1 - \Pr(N\geq 2E[N])\nonumber\\
& \geq & 1 - \frac{\mbox{Var}(N)}{E[N]^2}\nonumber\\
& = & 1-\frac{1}{E[N]}\nonumber\\
& = & 1-\frac{1}{\left(\frac{d(\lambda)}{2}+2\right) \left(\frac{3d(\lambda)}{2}+2\right)\lambda}.
\end{eqnarray}
By \eqref{d-epsilon}, \eqref{p-open} and \eqref{p-efficient-bound}, we have
\begin{equation}
p_o(d) \geq p_c(d)+p_e(d)-1 > 1-q_0.
\end{equation}

Now consider the {\em dual lattice} $\mathcal{L}'$ of $\mathcal{L}$. The construction of
$\mathcal{L}'$ is as follows: let each vertex of $\mathcal{L}'$ be located at the center of a
square of $\mathcal{L}$. Let each edge of $\mathcal{L}'$ be open if and only if it crosses an open
edge of $\mathcal{L}$, and closed otherwise. It is clear that each edge in $\mathcal{L}'$ is open
also with probability $p_o(d)$. Let
\[
q=1-p_o(d),
\]
and choose $2m$ edges in $\mathcal{L}'$. Because the states (i.e., open or closed) of any set of
non-adjacent edges are independent, we can choose $m$ edges among these $2m$ edges such that their
states are independent. As a result,
\[
\Pr(\mbox{All the $2m$ edges are closed})\leq q^m.
\]

\begin{figure}[t] \centering
\includegraphics[width=2.5in]{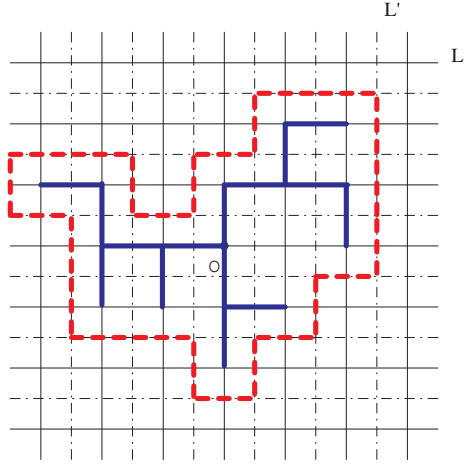}
\caption{If the origin belongs to an infinite open edge cluster in $\mathcal{L}$, then there
cannot exist a closed circuit surrounding the origin in $\mathcal{L}'$, and vice
versa.}\label{fig:DualLattice}
\end{figure}

Now a key observation is that if the origin belongs to an infinite open edge cluster in
$\mathcal{L}$, for which the event is denoted by $E_{\mathcal{L}}$, then there cannot exist a
closed circuit (a circuit consisting of closed edges) surrounding the origin in $\mathcal{L}'$,
for which the event is denoted by $E_{\mathcal{L}'}$, and vice versa. This is demonstrated in
Figure~\ref{fig:DualLattice}.
Hence
\[
\Pr(E_{\mathcal{L}})>0 \Longleftrightarrow \Pr(E_{\mathcal{L}'})<1.
\]
Furthermore, we have
\[
\Pr(E_{\mathcal{L}'}) = \sum_{m=2}^{\infty}\Pr(\exists \mathcal{O}_c(2m)) \leq \sum_{m=2}^{\infty}
\gamma(2m)q^m,
\]
where $\mathcal{O}_c(2m)$ is a closed circuit having length $2m$ surrounding the origin, and
$\gamma(2m)$ is the number of such circuits.

By Lemma~\ref{Lemma-Closed-Circuit-Number} in Appendix A, we have
\begin{eqnarray}
\sum_{m=2}^{\infty} \gamma(2m)q^m &\leq & \sum_{m=2}^{\infty} \frac{4}{27}(m-1)(9q)^m\nonumber\\
&= & \frac{4}{27}\left[\sum_{m=2}^{\infty} m(9q)^m-\sum_{m=2}^{\infty}(9q)^m\right]\nonumber\\
& =& \frac{4}{27}\left[\frac{2(9q)^2-(9q)^3}{(1-9q)^2}-\frac{(9q)^2}{1-9q}\right]\nonumber\\
& = &\frac{12q^2}{(1-9q)^2}.
\end{eqnarray}

Because
\[
q=1-p_o(d)<q_0=\frac{1}{9+2\sqrt{3}},
\]
we have $\frac{2\sqrt{3}q}{1-9q}<1$, and hence $\frac{12q^2}{(1-9q)^2}<1$. Thus the origin belongs
to an infinite open edge cluster in $\mathcal{L}$ with a positive probability. The existence of an
infinite open edge cluster in $\mathcal{L}$ implies the existence of an infinite connected
component in $G(\mathcal{H}_\lambda,1,q(\cdot))$, and this completes our proof for
Theorem~\ref{Theorem-General-Degree-Dependent}-(i). \qed

The first part of Theorem~\ref{Theorem-General-Degree-Dependent} provides a sufficient condition
for $G(\mathcal{H}_\lambda,1,q(\cdot))$ to have an infinite component. The second part of
Theorem~\ref{Theorem-General-Degree-Dependent} provides a sufficient condition for
$G(\mathcal{H}_\lambda,1,q(\cdot))$ to have no infinite component. Thus, it provides a {\em
necessary} condition for $G(\mathcal{H}_\lambda,1,q(\cdot))$ to have an infinite component. To
show this, we use another mapping between the continuum model and a discrete percolation model.

\begin{figure}[t!]
\centering
\includegraphics[width=4in]{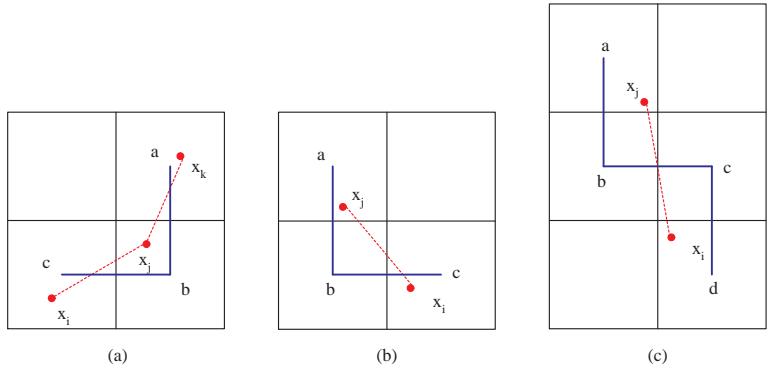}
\caption{Examples of open edges: in (a) and (b), $S_a$, $S_b$ and $S_c$ are open, while in (c),
$S_a$, $S_b$, $S_c$ and $S_d$ are all open.}\label{fig:OpenSite}
\end{figure}

\vspace{+.1in}%
\emph{Proof of Theorem \ref{Theorem-General-Degree-Dependent}-(ii):} Map
$G(\mathcal{H}_\lambda,1)$ to a square lattice $\mathcal{L}$ with edge length
$d=\frac{\sqrt{2}}{2}$. Let the square centered at vertex $a$ with edge length $d$ be $S_a$. Let
$N(S_a)$ and $N'(S_a)$ be the number of nodes of $G(\mathcal{H}_\lambda,1)$ and
$G(\mathcal{H}_\lambda,1, q(\cdot))$ in $S_a$, respectively. We say $S_a$ is \emph{open} if and
only if either one of the following conditions holds:
\begin{itemize}
\item[(i)] $N'(S_a)\geq 1$; \item[(ii)] There is a link of $G(\mathcal{H}_\lambda,1,q(\cdot))$
crossing $S_a$ which directly connects two nodes of $G(\mathcal{H}_\lambda,1,q(\cdot))$ outside
$S_a$.
\end{itemize}
In Figure~\ref{fig:OpenSite}, we illustrate the possible examples of open squares in
$\mathcal{L}$. If $S_a$ is open only because $S_a$ satisfies condition (ii), we say it is {\em
type-2 open}; otherwise, we say it is {\em type-1 open}.

The probability that $S_a$ is type-1 open can be expressed as
\begin{eqnarray}\label{p-1}
p_1 & = & \Pr(N'(S_a)\geq 1)\nonumber\\
& = & \sum_{k=1}^{\infty} \Pr(N(S_a)=k, N'(S_a)\geq 1)\nonumber\\
& = & \sum_{k=1}^{\infty} \Pr(N(S_a)=k)\Pr(N'(S_a)\geq 1|N(S_a)=k).
\end{eqnarray}
When $q(k)$ is non-decreasing in $k$, by Appendix B,
\begin{eqnarray}\label{p-1-bound-1}
p_1 \leq  1-e^{-\frac{\lambda}{2}}-\sum_{k=1}^{\infty}
\frac{(\frac{\lambda}{2})^k}{k!}e^{-\frac{\lambda}{2}}q(k-1)^k.
\end{eqnarray}
If~\eqref{qk-lower-bound-1} holds, we have $p_1<\frac{1}{27}$. When $q(k)$ is non-increasing in
$k$, by Appendix C,
\begin{eqnarray}\label{p-1-bound-2}
p_1 \leq \sum_{k=1}^{\infty}\frac{\left(\frac{\lambda}{2}\right)^k}{k!}e^{-\frac{\lambda}{2}}
\sum_{m=0}^{\infty}\frac{[\lambda(2\sqrt{2}+\pi)]^m}{m!}e^{-\lambda(2\sqrt{2}+\pi)}
\left(1-q(m+k-1)^k\right).
\end{eqnarray}
If~\eqref{qk-lower-bound-2} holds, we have $p_1<\frac{1}{27}$ as well. Therefore, in both cases,
we have $p_1<\frac{1}{27}$.

\begin{figure}[t!]
\centering
\includegraphics[width=2.5in]{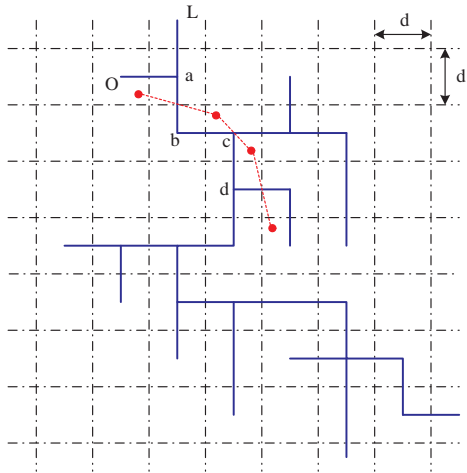}
\caption{An open path in lattice $\mathcal{L}$}\label{fig:OpenPath}
\end{figure}

If there is an infinite component in $G(\mathcal{H}_\lambda,1,q(\cdot))$, there must exist an
infinite path consisting of nodes in $G(\mathcal{H}_\lambda,1,q(\cdot))$. Furthermore, this
infinite path must pass through an infinite number of open squares in $\mathcal{L}$, as
illustrated in Figure~\ref{fig:OpenPath}. This is because along the infinite path in
$G(\mathcal{H}_\lambda,1,q(\cdot))$, each square of $\mathcal{L}$ contains at least one node of
$G(\mathcal{H}_\lambda,1,q(\cdot))$ or is crossed by a link of $G(\mathcal{H}_\lambda,1,q(\cdot))$
that directly connects two nodes of $G(\mathcal{H}_\lambda,1,q(\cdot))$ outside $S_a$.

Now choose a path in $\mathcal{L}$ starting from the origin\footnote{Note that the choice of the
origin is arbitrary.} having length $3m$. From Figure~\ref{fig:OpenSite}, we can see that a link
from any given node in $G(\mathcal{H}_\lambda,1,q(\cdot))$ can go through at most three open
squares in addition to the open square containing the given node.  As a result, along the path,
among every three consecutive open squares, there exists at least one type-1 open square. Thus, we
have
\begin{equation}
\Pr(\mbox{All the $3m$ edges are open})\leq p_1^{m+1}.
\end{equation}
Now
\begin{equation}
\Pr(\exists \mathcal{O}_p(3m))\leq \xi(3m)p_1^{m+1},
\end{equation}
where $\mathcal{O}_p(3m)$ is an open path in $\mathcal{L}$ starting from the origin with length
$3m$, and $\xi(3m)$ is the number of such paths. For a path in $\mathcal{L}$ from the origin, the
first edge has four choices for its direction, and all other edges have at most three choices for
their directions. Therefore, we have
\begin{equation}\label{xi-m}
\xi(3m)\leq 4\cdot 3^{3m-1},
\end{equation}
and
\begin{equation}\label{open-path-bound}
\Pr(\exists \mathcal{O}_p(3m))\leq 4\cdot3^{3m-1}p_1^{m+1}=\frac{4}{3}p_1(3^3p_1)^m.
\end{equation}
When $p_1<\frac{1}{27}$, the RHS of (\ref{open-path-bound}) converges to 0 as $m \rightarrow \infty$. This implies that
with probability 1, there is no infinite path starting from the origin (which is arbitrary) in $\mathcal{L}$.
Therefore, with probability 1, there is no infinite component in $G(\mathcal{H}_\lambda,1,q(\cdot))$ either.\qed

As an example of degree-dependent failures one may consider a strategy where an attacker sets a threshold $\phi$ and
destroys all nodes having degree strictly greater than $\phi$. Given $G(\mathcal{H}_{\lambda},1)$ and an integer
$\phi$, all nodes with degree strictly greater than $\phi$ and their associated links fail, and all other nodes remain
operational. That is
\begin{equation}\label{eq:q-k}
q(k)=\left\{\begin{array}{ll} 0 & k\leq \phi\\ 1 &  k\geq\phi+1 \end{array}\right.
\end{equation}
Let the remaining graph be denoted by $G(\mathcal{H}_{\lambda},1,\phi)$. By directly applying Theorem
\ref{Theorem-General-Degree-Dependent}-(i), we know that there exists $k_1<\infty$, such that when $\phi \geq k_1$,
$G(\mathcal{H}_{\lambda},1,\phi)$ is percolated.

\begin{figure}[t!]
\centering
\includegraphics[width=3in]{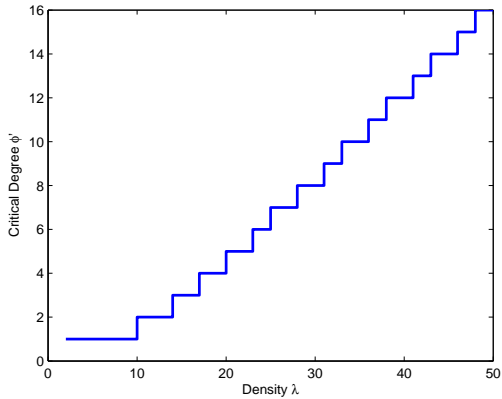}
\caption{Critical value of $\phi$ obtained by
condition~\eqref{eq:phi-prime-result-2}}\label{fig:CriticalDegree}\vspace{-.2in}
\end{figure}

We can also apply Theorem \ref{Theorem-General-Degree-Dependent}-(ii) to obtain a lower bound on the critical value of
$\phi$. By substituting~\eqref{eq:q-k} into~\eqref{qk-lower-bound-1}, we see that if $\phi'$ satisfies
\begin{equation}\label{eq:phi-prime}
e^{-\frac{\lambda}{2}}+\sum_{k=\phi'+2}^{\infty}\frac{\left(\frac{\lambda}{2}\right)^k}{k!}e^{-\frac{\lambda}{2}}
>1-{\frac{1}{27}},
\end{equation}
then for any $\phi \leq \phi'$, $G(\mathcal{H}_{\lambda},1,\phi)$ is not percolated.
Condition~\eqref{eq:phi-prime} can be simplified as
\begin{equation}\label{eq:phi-prime-result-2}
\sum_{k=0}^{\phi'+1}\frac{\left(\frac{\lambda}{2}\right)^k}{k!}
<{\frac{1}{27}}e^{\frac{\lambda}{2}}+1.
\end{equation}
For any given $\lambda$, we can use~\eqref{eq:phi-prime-result-2} to find the critical value of
$\phi$. Figure~\ref{fig:CriticalDegree} plots the maximal $\phi'$ against $\lambda$
satisfying~\eqref{eq:phi-prime-result-2}.

\section{Cascading Node Failures}

As we pointed out before, in networks which carry load, distribute a resource or aggregate data,
such as wireless sensor networks and electrical power networks, the failure of one node often
results in the redistribution of the load from the failed node to other nearby nodes. If nodes
fail when the load on them exceeds some maximum capacity or when the battery energy is depleted,
then a {\em cascading failure} or avalanche may occur because the redistribution of the load
causes other nodes to exceed their thresholds and fail, thereby leading to a further
redistribution of the load.

Cascades have been used in social networks to model phenomena such as epidemic spreading, belief
propagation, etc. Although they are generated by different mechanisms, cascades in social and
economic systems are similar to a cascading failure in physical infrastructure
networks~\cite{KoTaMi99, SaCaLy00} in that initial failures can increase the likelihood of
subsequent failures, leading to eventual dramatic global outages. Usually, such cascading failures
are extremely difficult to predict, even when the properties of individual components are well
understood. In~\cite{Wa02, Ne03}, the author investigate such cascading failures in social
networks by modelling the problem as a binary decision percolation process on random networks
where the links between distinct pairs of nodes are independent.

In contrast to previous work, we study cascading failures in large-scale wireless networks modelled by random geometric
graphs. To our knowledge, this is the first investigation of cascading phenomena in networks with geometric
constraints. In particular, we consider the following model. Consider a network modelled by a random geometric graph
$G(\mathcal{H}_{\lambda},1)$ with $\lambda>\lambda_c$, where an initial failure seed is represented by a single failed
node. This initial failure seed is an exogenous event (shock) that is very small relative to the whole network. We are
interested in whether this initial small shock can lead to a global cascade of failures, which is technically defined
as follows.

Note that in characterizing cascading failure, the essential point is to assess whether the network has been affected
in a global manner, rather than in an isolated local manner.  For this reason, cascades cannot be easily characterized
by, for instance, what percentage of the network nodes have failed.  Instead, after some thought, one is led to the
conclusion that percolation (the existence of an infinite failed component) is an appropriate notion with which to
characterize cascading failures. Thus, we have the following definition.

\vspace{+.1in}
\begin{definition}
A cascading failure is an ordered sequence of node failures triggered by an initial failure seed
resulting in an infinite component of failed nodes in the network.
\end{definition}
\vspace{+.1in}

To describe cascading failures, we use the following simple but descriptive model. We assume that
due to redistribution of the load, each node $i$ fails if a given fraction $\psi_i$ of its
neighbors have failed, where the $\psi_i$'s are i.i.d. random variables with probability density
function $f(\psi)$. The order of the failure sequence is then the topological order determined by
the location of the initial failure and the threshold $\psi_i$ of each node $i$.

\begin{figure}[t!]
\centerline{ \subfigure[]{
\includegraphics[width=1.2in]{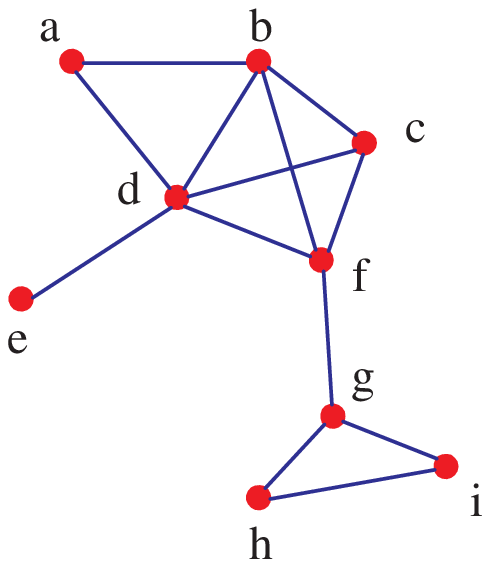}}\hfil
\subfigure[]{
\includegraphics[width=1.2in]{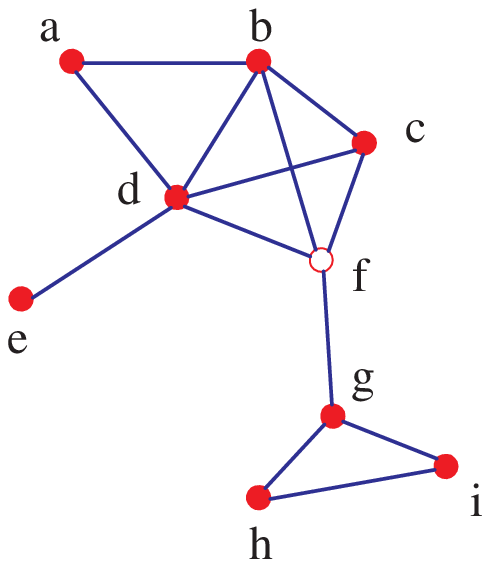}}\hfil
\subfigure[]{
\includegraphics[width=1.2in]{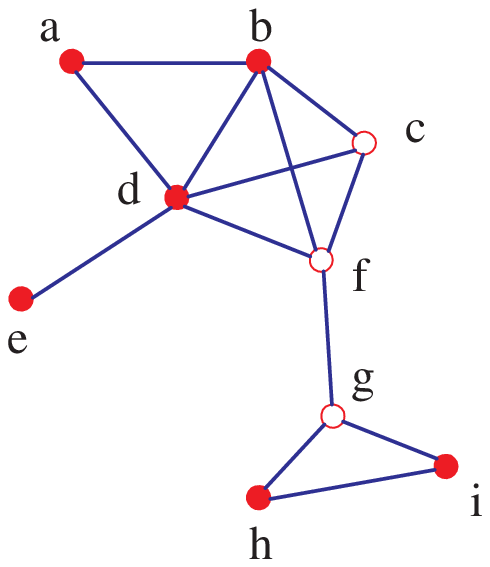}}}
\centerline{ \subfigure[]{
\includegraphics[width=1.2in]{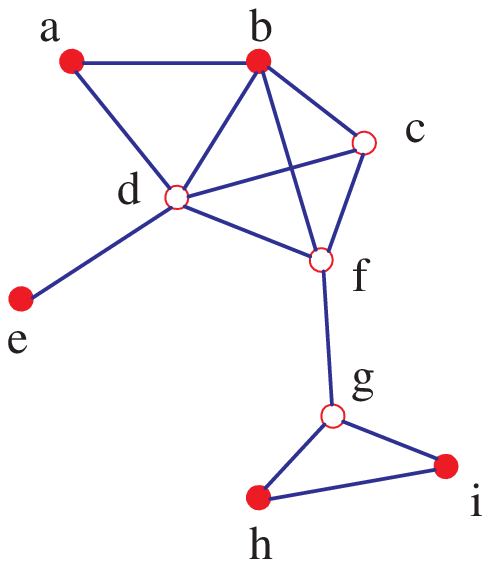}}\hfil
\subfigure[]{
\includegraphics[width=1.2in]{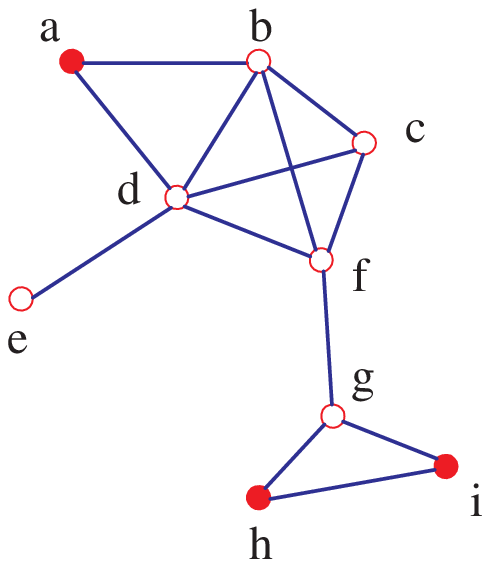}}\hfil
\subfigure[]{
\includegraphics[width=1.2in]{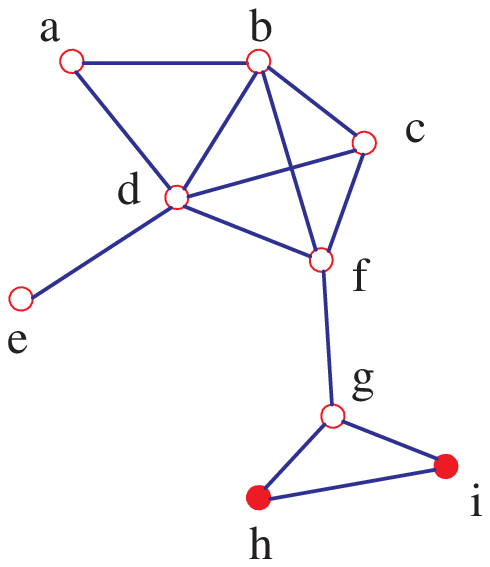}}}
\caption{An example of cascading failures (solid nodes are operational and empty circles are
failed)}\label{fig:Cascading}
\end{figure}

Unlike the degree-dependent scenarios studied earlier, cascading failure processes exhibit dynamic
evolution. This is illustrated in Figure~\ref{fig:Cascading}. The simple network in
Figure~\ref{fig:Cascading}-(a) has nine nodes: $a$, $b, ... , i$ with failure thresholds
$\psi_a=0.8, \psi_b=0.7, \psi_c=0.1, \psi_d=0.3, \psi_e=0.4, \psi_f=0.5, \psi_g=0.2, \psi_h=0.6,
\psi_i=0.9$.
At the beginning, as shown in Figure~\ref{fig:Cascading}-(b), an initial failure occurs at node
$f$. Then, since $\psi_c=0.1$ and one of $c$'s three neighbors has failed, node $c$ fails.
Similarly, node $g$ also fails. This is illustrated in Figure~\ref{fig:Cascading}-(c). Since
$\psi_d=0.3$, $d$ does not fail until two of its five neighbors have failed
(Figure~\ref{fig:Cascading}-(d)). This process continues (Figure~\ref{fig:Cascading}-(e)) until no
further failures can occur in the network (Figure~\ref{fig:Cascading}-(f)). The resulting network
is denoted by $G(\mathcal{H}_{\lambda},1,\psi)$, which, in this example, has failed nodes $a$, $b,
... , g$, and operational nodes $h$ and $i$. The ordered sequence of failures in this example is
$f, \{c,g\}, d, \{b, e\}, a$.

For two \emph{adjacent} nodes $u$ and $v$, we say that node $u$'s failure is \emph{caused} by node
$v$'s failure if and only if node $u$'s failure immediately follows node $v$'s failure in the
ordered failure sequence. In the example of Figure~\ref{fig:Cascading}, node $c$'s failure is
caused by node $f$'s failure, and node $d$'s failure is caused by node $c$'s failure.

Now observe that the initial failure can grow only when some neighbor, say $j$, of the initial
failure seed has a threshold satisfying $\psi_j\leq \frac{1}{k_j}$, where $k_j\geq 1$ is the
degree of $j$. We call such a node \emph{vulnerable}. The probability of a node being vulnerable
is
\begin{equation}\label{rho-k}
\rho_k = F_{\psi}\left(\frac{1}{k}\right)=\int_{0}^{\frac{1}{k}} f(\psi)d\psi,
\end{equation}
where $F_{\psi}(\cdot)$ is the cumulative distribution function of $\psi_j$. In the example of
Figure~\ref{fig:Cascading}, nodes $c$, $e$ and $g$ are vulnerable.

When the initial failure seed is directly connected to a component of vulnerable nodes, all nodes in this component
fail. The extent of the failure, and hence the resilience of the network, depends not only on the number of vulnerable
nodes, but also on how they are connected to one another. In the context of this model, a cascade of failed nodes forms
when the network has an infinite component of vulnerable nodes and the initial failure seed is either inside this
component or adjacent to some node in this component.

On the other hand, if node $i$ has a threshold satisfying $\psi_i> \frac{k_i-1}{k_i}$, where $k_i$
is the degree of $i$, then node $i$ will not fail as long as at least one neighbor is operational.
We call such a node \emph{reliable}. Otherwise, if $\psi_i \leq \frac{k_i-1}{k_i}$,  we call node
$i$ \emph{unreliable}. For $k\geq 1$, the probability of a node being reliable is given by
\begin{equation}\label{sigma-k}
\sigma_k = 1-F_{\psi}\left(\frac{k-1}{k}\right)=\int_{\frac{k-1}{k}}^1 f(\psi)d\psi.
\end{equation}
For $k=0$, we set $\sigma_0=1$. Intuitively, a node $i$ with no neighbors should be reliable,
since it remains operational no matter what $\psi_i$ is, unless node $i$ itself is the initial
failure. This also agrees with~\eqref{sigma-k} by applying the convention $F(-\infty)=0$. In the
example of Figure~\ref{fig:Cascading}, nodes $a$, $h$ and $i$ are reliable, and all the other
nodes are unreliable. Note that when two reliable nodes are adjacent and neither is an initial
failure seed, no matter what else happens in the network, they remain operational. This is
illustrated by nodes $h$ and $i$ in Figure~\ref{fig:Cascading}. When a reliable node $u$ has only
unreliable neighbors, node $u$ fails if and only if all its unreliable neighbors fail, unless node
$u$ is the initial failure. We call such a reliable node an \emph{isolated reliable} node.

The following theorem presents our main results on cascading failures in wireless networks. It
provides a sufficient condition for the existence of an infinite component of vulnerable nodes, as
well as a sufficient condition for the non-existence of an infinite component of unreliable nodes.
The theorem asserts that when there exists an infinite component of vulnerable nodes and the
initial failure is either inside this component or adjacent to some node in this component, then
there is a cascading failure in $G(\mathcal{H}_{\lambda},1)$. On the other hand, when there is no
infinite component of unreliable nodes, then there is no cascading failure no matter where the
initial failure is.

\vspace{0.1in}%
\begin{theorem}\label{Corollary-Cascading-Failures} (i) For any $\mu_1>\mu_c$ and $G(\mathcal{H}_{\lambda},1)$ with $\mu>\mu_1$,
there exists
$k_0<\infty$ depending on $\mu$ such that if
\begin{equation}\label{eq:F-psi}
F_{\psi}\left(\frac{1}{k_0}\right) \geq\frac{\mu_1}{\mu},
\end{equation}
then with probability 1, there exists an infinite component of vulnerable nodes in
$G(\mathcal{H}_{\lambda},1)$. Moreover, if the initial failure is inside this component or
adjacent to some node in this component, then with probability 1, there is a cascading failure in
$G(\mathcal{H}_{\lambda},1)$.

(ii) For any $G(\mathcal{H}_{\lambda},1)$ with $\mu>\mu_c$, if
\begin{equation}\label{sigma-upper-bound}
\sum_{k=1}^{\infty}\frac{\left(\frac{\lambda}{2}\right)^k}{k!}e^{-\frac{\lambda}{2}}
\sum_{m=0}^{\infty}\frac{[\lambda (2\sqrt{2}+\pi)]^m}{m!}e^{-\lambda
(2\sqrt{2}+\pi)}\left(1-\left[1-F_{\psi}\left(\frac{m+k-2}{m+k-1}\right)\right]^k\right)<\frac{1}{27},
\end{equation}
where $F_{\psi}(-\infty)=0$ by convention, then with probability 1, there is no infinite component
of unreliable nodes. As a consequence, with probability 1, there is no cascading failure in
$G(\mathcal{H}_{\lambda},1)$ no matter where the initial failure is.
\end{theorem}
\vspace{+.1in}%

\emph{Proof:} To prove (i), we view the problem as a degree-dependent node failure problem where a
vulnerable node is considered ``operational" and a non-vulnerable node is considered a ``failure."
In this model, each node with degree $k$ fails with a probability $1-\rho_k$.  Then, by applying
Theorem \ref{Theorem-General-Degree-Dependent}-(i) directly, we have for any $\mu_1>\mu_c$ and
 $G(\mathcal{H}_{\lambda},1)$ with $\mu>\mu_1$, there
exists $k_0<\infty$ such that if
\begin{equation}\label{rhok-upper-bound}
\rho_{k_0} =F_{\psi}\left(\frac{1}{k_0}\right) \geq\frac{\mu_1}{\mu},
\end{equation}
then with probability 1, there exists an infinite component of vulnerable nodes in
$G(\mathcal{H}_{\lambda},1)$. If the initial failure is inside this component or adjacent to some
node in this component, then there is a cascading failure in $G(\mathcal{H}_{\lambda},1)$.

To prove (ii), we first show (a): if~\eqref{sigma-upper-bound} holds, then with probability 1,
there is no infinite component of unreliable nodes. We then show (b): if there is no infinite
component of unreliable nodes, then with probability 1, there is no cascading failure no matter
where the initial failure is.

To show (a), we apply the result of Theorem~\ref{Theorem-General-Degree-Dependent}-(ii). Regard an
unreliable node as ``operational" and a reliable node as a ``failure". Then, $\sigma_k$---the
probability of a node with degree $k$ being reliable---becomes the failure probability $q(k)$ in
the context of Theorem~\ref{Theorem-General-Degree-Dependent}-(ii). Since $\sigma_k$ is
non-increasing in $k$, we replace $q(m+k-1)$ in~\eqref{qk-lower-bound-2} with
$\sigma_{m+k-1}=1-F_{\psi}\left(\frac{m+k-2}{m+k-1}\right)$ and obtain~\eqref{sigma-upper-bound}.
By Theorem~\ref{Theorem-General-Degree-Dependent}-(ii), when~\eqref{sigma-upper-bound} holds, with
probability 1, there is no infinite component of unreliable nodes in the network.

In order to show (b), we will show that if there is a cascading failure, i.e., there is an
infinite component $W$ of failed nodes, there must exist an infinite component of unreliable nodes
in the network. Assume the initial failure takes place at node $u$, and consider two cases: (1)
node $u$ is an unreliable node or an isolated reliable node; (2) node $u$ is a non-isolated
reliable node.

For case (1), if there is an infinite component of failed nodes in the network, all the failed
nodes are either unreliable or isolated reliable. This is because non-isolated reliable nodes do
not fail no matter what happens in the network. Furthermore, except for the initial failure, an
isolated reliable node fails if and only if all its (unreliable) neighbors fail. This implies that
except for the initial failure, the failure of any isolated reliable node does not cause any other
failures. In other words, except for the initial failure, the failure of any unreliable node is caused by the failure(s) of other
unreliable node(s). Thus, all the unreliable nodes in $W$ belong to the same component.

\begin{figure}[t!]
\centering
\includegraphics[width=1.5in]{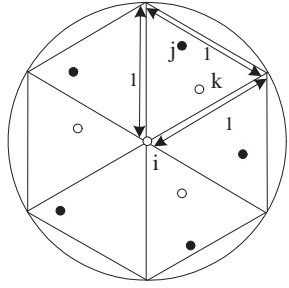}
\caption{For any node $i$, the number of its isolated reliable neighbors cannot be strictly
greater than 6. This is because any two nodes $j$ and $k$ inside one of six fan-shaped regions are
adjacent to each other. Thus, nodes $j$ and $k$ cannot be isolated reliable by definition. By the
same argument, when $i$ is the initial failure, the number of induced isolated reliable nodes
cannot be strictly greater than 6.}\label{fig:Neighbors}
\end{figure}

Now suppose there is only a finite number of unreliable nodes in $W$. Then there must be an
infinite number of isolated reliable nodes in $W$. Note first that an isolated reliable node
cannot be adjacent to another isolated reliable node by definition. Furthermore, as illustrated in
Figure~\ref{fig:Neighbors}, each unreliable node cannot have strictly more than 6 isolated
reliable neighbors.  Therefore, it is impossible to have a finite number of unreliable nodes but
an infinite number of isolated reliable nodes in $W$. This contradiction ensures that the
component of unreliable nodes in $W$ is infinite.

For case (2), a non-isolated reliable node fails if and only if (i) it is adjacent to the initial
failure, (ii) not adjacent to any other reliable nodes, and (iii) all of its unreliable neighbors
fail. We call a non-isolated reliable node satisfying condition (i)--(ii) an \emph{induced
isolated reliable} node. As illustrated in Figure~\ref{fig:Neighbors}, the number of induced
isolated reliable nodes cannot be strictly greater than 6. Except for the initial failure and a
finite number of induced isolated reliable nodes, all other failed nodes in $W$ are either
isolated reliable or unreliable. Observe that as in the failure of an isolated reliable node, the
failure of an induced isolated reliable node does not cause any other failures. In other words, except for the initial failure,
the failure of any unreliable node is caused by the failure(s) of other unreliable node(s). Thus,
all the unreliable nodes in $W$ belong to the same component. Then by the same argument for the
first case, there must exist an infinite component of unreliable nodes in $W$. \qed

\section{Simulation Studies}

\begin{figure}[t!]
\centerline{ \subfigure[$q(k)=\max\{0,1-\frac{\mu_c}{\mu}-\frac{1}{k}\}$; solid: operational nodes, empty: failed
nodes]{
\includegraphics[width=2.8in]{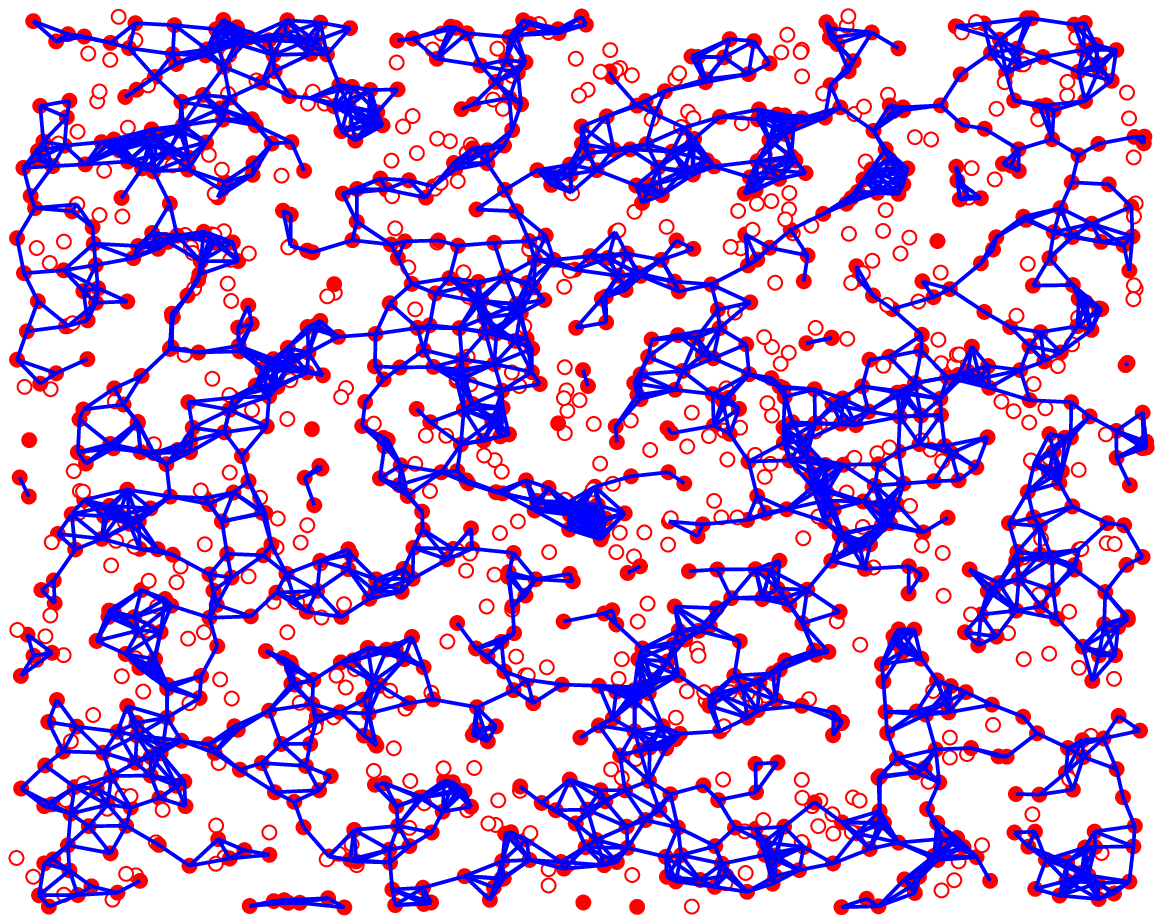}}\hfil
\subfigure[$q(k)=0,k\leq 4$, and $q(k)=1, k>4$; solid: operational nodes, empty: failed nodes]{
\includegraphics[width=2.8in]{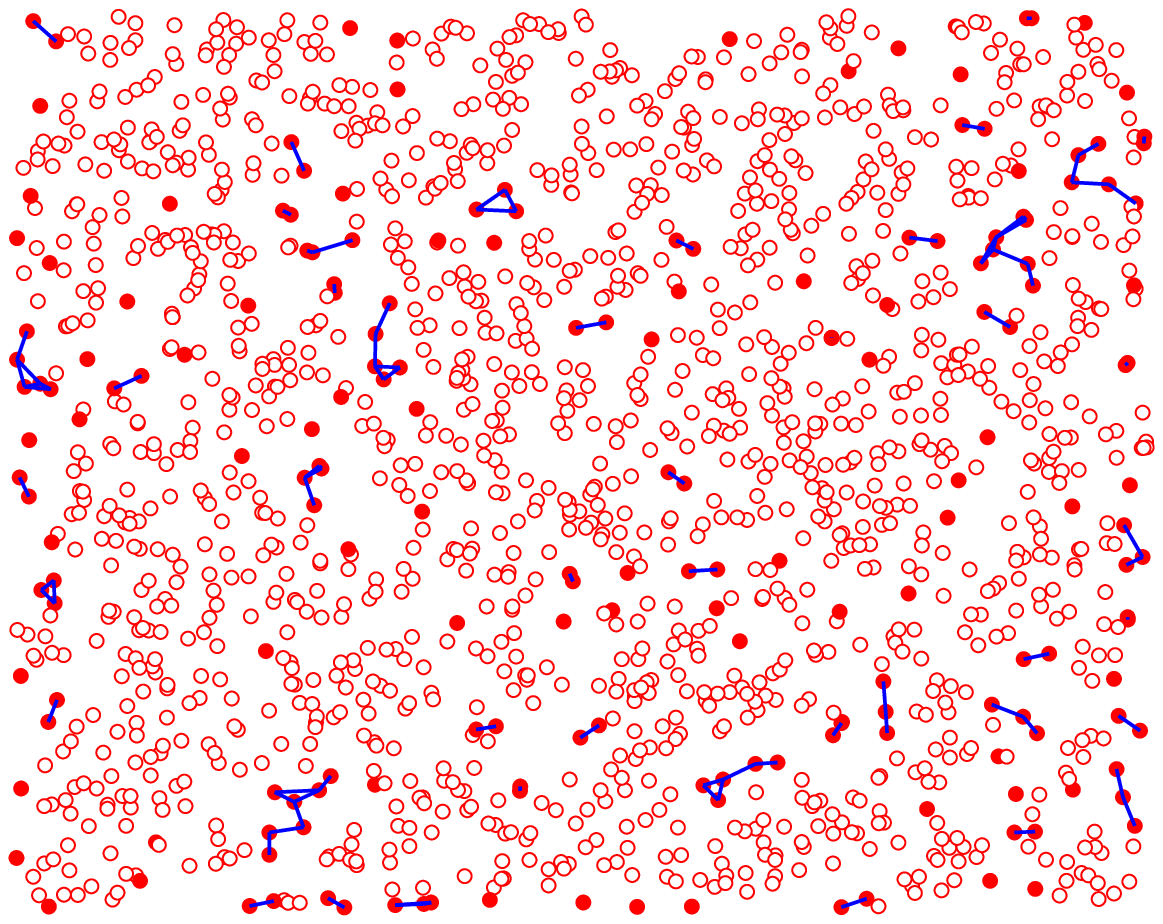}}}
\caption{Degree-dependent node failures in $G(\mathcal{H}_{\lambda},1)$ with
$\mu=8$.}\label{fig:RGGDegreeDependent1}
\end{figure}

We illustrate degree-dependent node failures with two examples in Figure~\ref{fig:RGGDegreeDependent1}. The original
network has $n=1600$ nodes uniformly distributed in $[0,25]^2$, and mean degree $\mu=8.04$. In
Figure~\ref{fig:RGGDegreeDependent1}-(a), $q(k)=\max\{0,1-\frac{\mu_c}{\mu}-\frac{1}{k}\}$. This function satisfies
condition (\ref{qk-upper-bound}) and the remaining network of operational nodes still has a large connected component
spanning almost the whole network, where empty circles represent failed nodes. In
Figure~\ref{fig:RGGDegreeDependent1}-(b), $q(k)=0,k\leq 4$, and $q(k)=1, k>4$. This function satisfies condition
(\ref{qk-lower-bound-1}) and the remaining network of operational nodes consists of small isolated components.

\begin{figure}[t!]
\centerline{ \subfigure[original network]{
\includegraphics[width=2.8in]{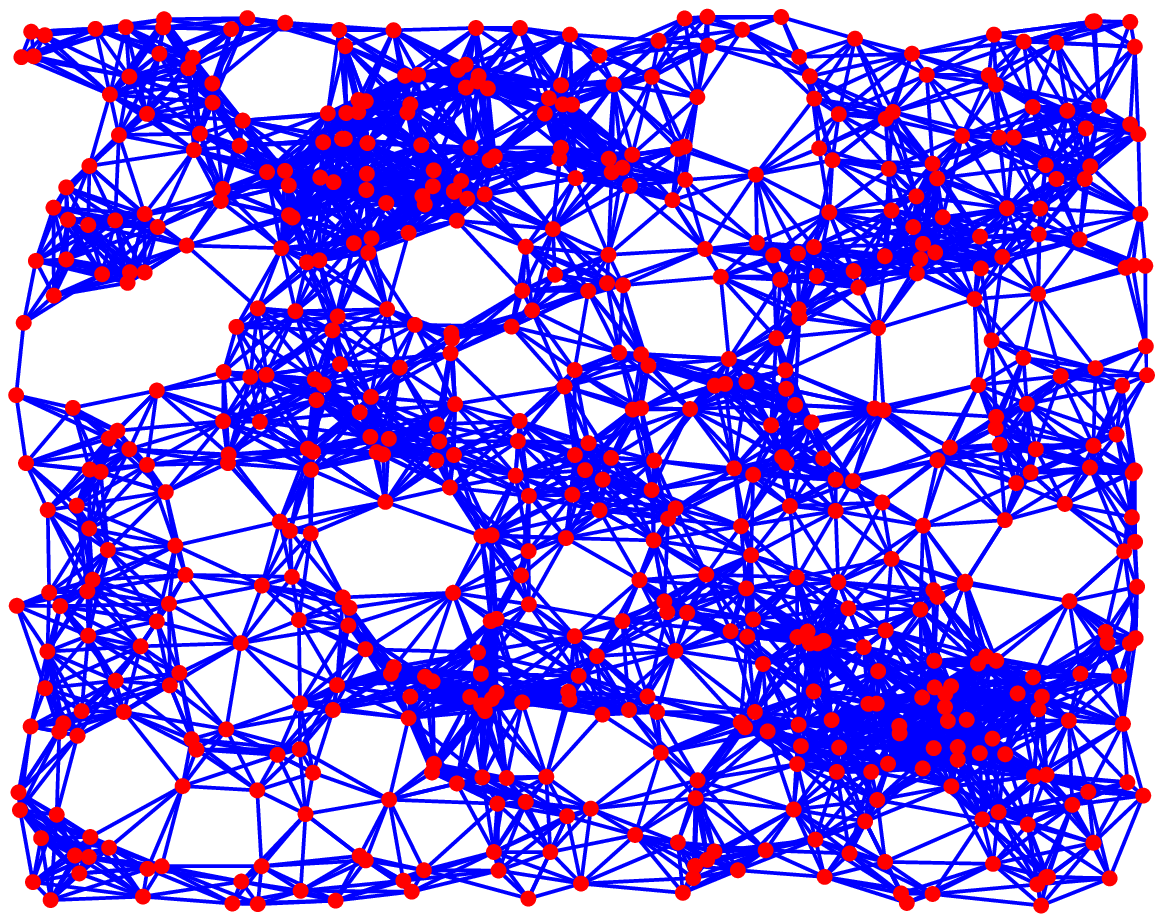}}\hfil
\subfigure[large component of vulnerable nodes; solid: non-vulnerable nodes, empty: vulnerable
nodes]{
\includegraphics[width=2.8in]{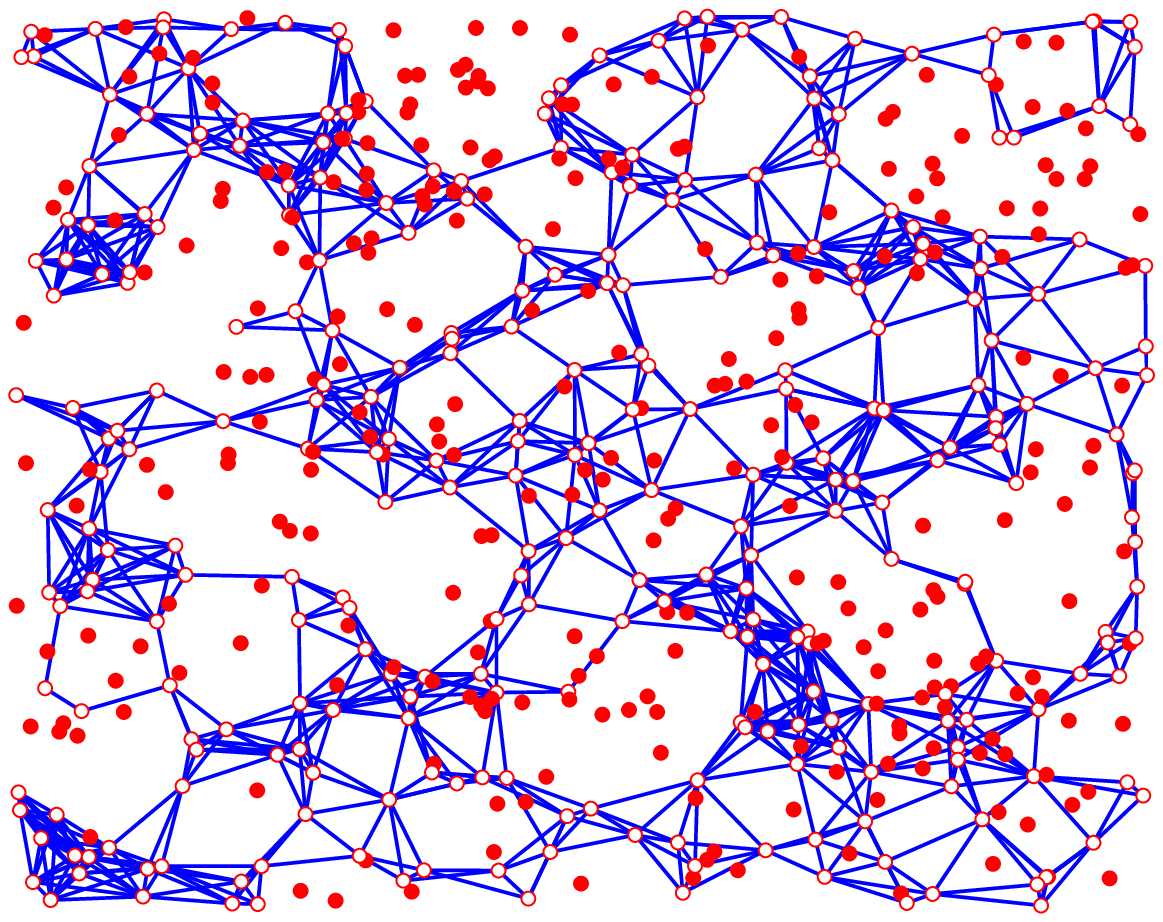}}}
\vspace{+0.05in}%
\centerline{ \subfigure[initial failure pointed to by arrow]{
\includegraphics[width=2.8in]{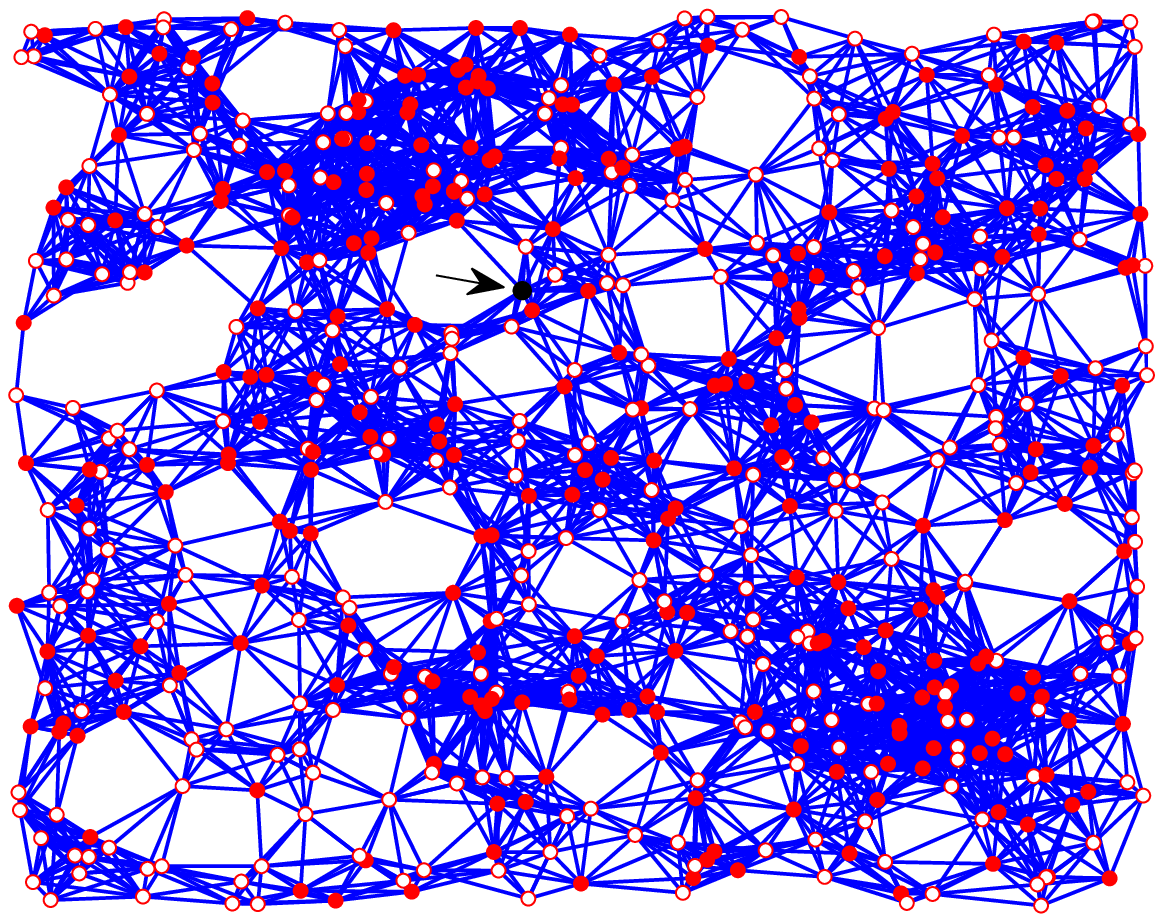}}\hfil
\subfigure[cascading failure; solid: operational nodes, empty: failed nodes]{
\includegraphics[width=2.8in]{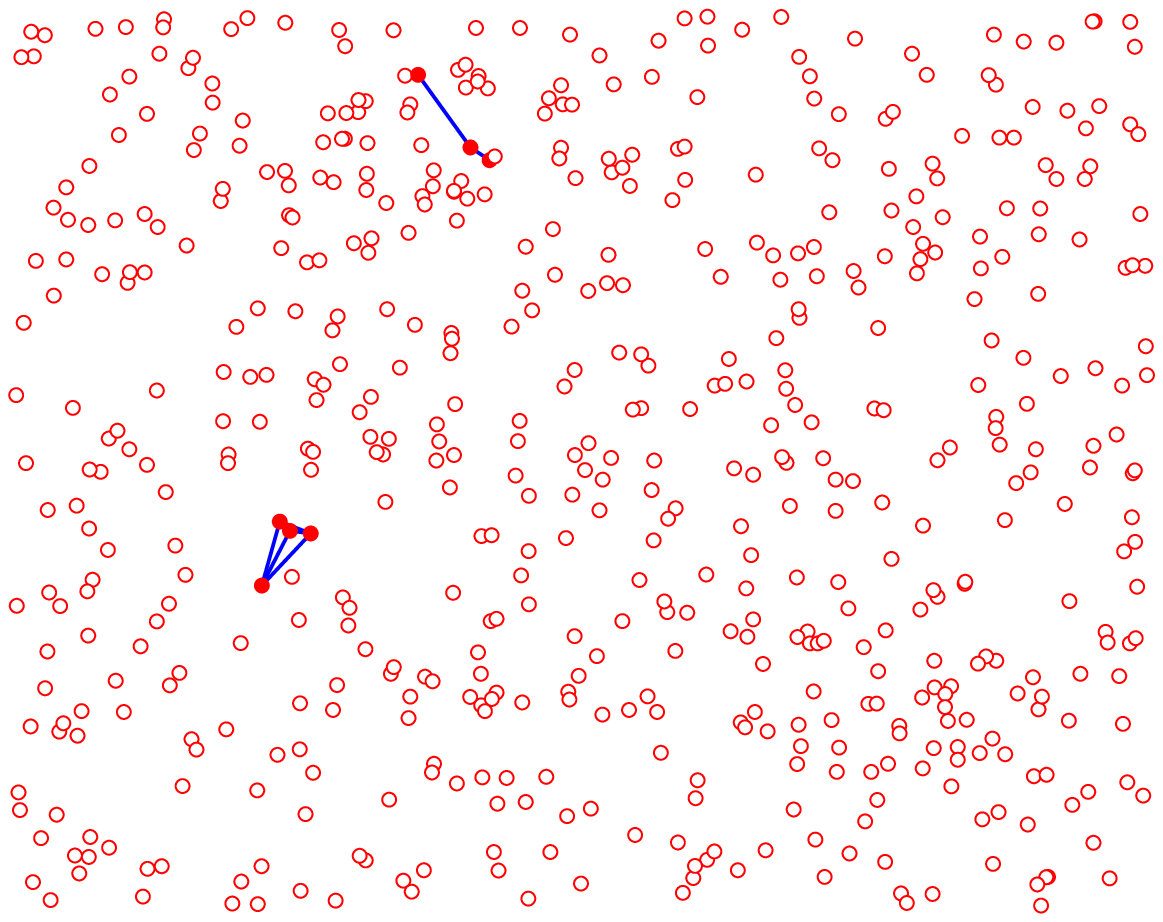}}}
\caption{Cascading failure} \label{fig:CascadingSim}
\end{figure}

In Figure~\ref{fig:CascadingSim}, we illustrate cascading failures. Figure~\ref{fig:CascadingSim}-(a) depicts a
wireless network with $n=1600$ nodes uniformly distributed in $[0,15]^2$, and mean degree $\mu=19.64$. Each node $i$
has a probability density function $f(\psi_i)$ for the threshold $\psi_i$, where $f(\psi_i)=\frac{15}{2}$ for
$0<\psi_i\leq 0.1$, and $f(\psi_i)=\frac{5}{18}$ for $0.1<\psi_i<1$. The $\psi_i$'s are assumed to be i.i.d.
Figure~\ref{fig:CascadingSim}-(b) depicts the largest component of vulnerable nodes (which are represented by empty
circles) spanning the network. Figure~\ref{fig:CascadingSim}-(c) indicates an initial failure caused by exogenous
event, which is represented by a black solid node pointed to by an arrow. From Figure~\ref{fig:CascadingSim}-(d), we
see that the resulting network suffers from a cascading failure, where the failed nodes are represented by empty
circles.

\begin{figure}[t!]
\centerline{ \subfigure[no large component of unreliable nodes; solid: reliable nodes, empty:
unreliable nodes]{
\includegraphics[width=2.8in]{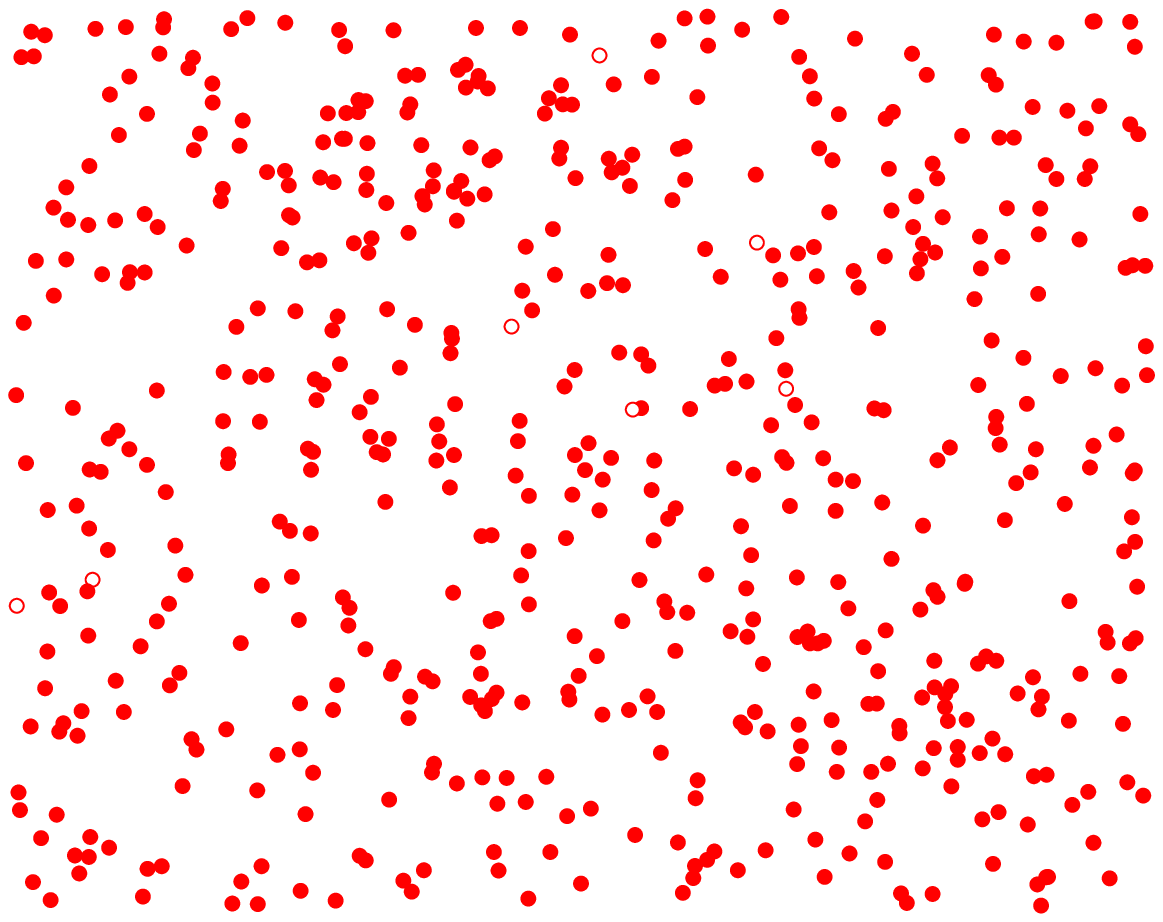}}\hfil
\subfigure[no cascading failure; solid: operation nodes, empty: failed nodes]{
\includegraphics[width=2.8in]{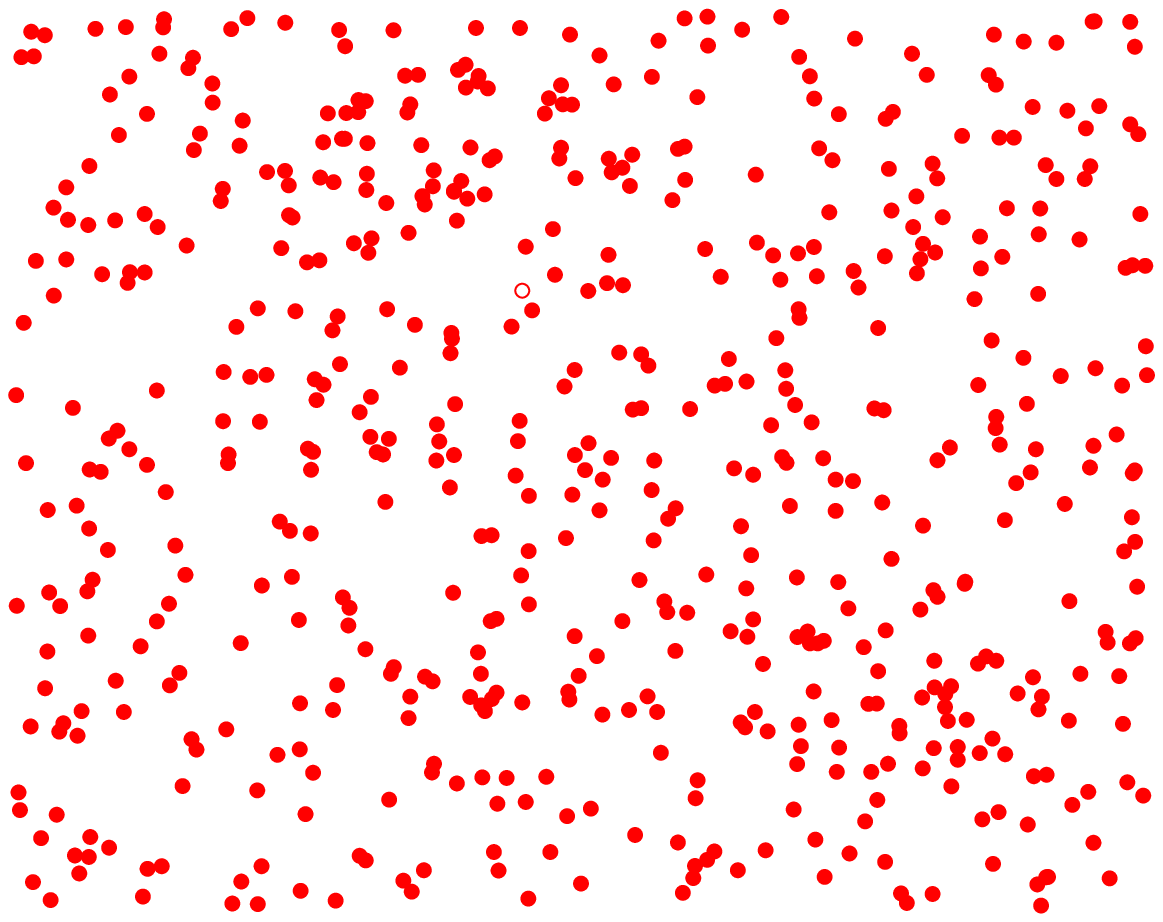}}}
\caption{No cascading failure}\label{fig:NoCascadingSim}
\end{figure}

Figure~\ref{fig:NoCascadingSim} illustrates an example where no cascading failure occurs. The network is the same as
the one shown in Figure~\ref{fig:CascadingSim}-(a). Here, each node $i$ has a probability density function $f(\psi_i)$
for the threshold $\psi_i$, where $f(\psi_i)=\frac{1}{999}$ for $0<\psi_i\leq 0.999$, and $f(\psi_i)=999$ for
$0.999<\psi_i<1$. The $\psi_i$'s are assumed to be i.i.d. This function satisfies the
condition~\eqref{sigma-upper-bound}. Figure~\ref{fig:NoCascadingSim}-(a) shows that there is no large component of
unreliable nodes (which are represented by empty circles) spanning the network. After the same initial failure as shown
in Figure~\ref{fig:CascadingSim}-(c) takes place, we see from Figure~\ref{fig:NoCascadingSim}-(b) that the initial
failure cause no other failures (failed nodes are represented by empty circles), and no cascading failure occurs in the
network.

\section{Conclusion}

In this paper, we studied network resilience problems from a percolation-based perspective. To analyze realistic
situations where the failure probability of a node depends on its degree, we introduced the degree-dependent failures
problem. We model this phenomenon as a degree-dependent site percolation process on random geometric graphs. Due to its
non-Poisson structure, degree-dependent site percolation is far from a trivial generalization of independent site
percolation. Using coupling methods and renormalization arguments, we obtained analytical conditions for the occurrence
of phase transitions within this model. Furthermore, in networks carrying traffic load, such as wireless sensor
networks and electrical power networks, the failure of one node can result in redistribution of the load onto other
nearby nodes. If these nodes fail due to excessive load, then this process can result in cascading failures. We
analyzed this cascading failure problem in large-scale wireless networks, and showed that it is equivalent to a
degree-dependent percolation process on random geometric graphs. We obtained analytical conditions for the occurrence
and non-occurrence of cascading failures, respectively. To our knowledge, this work represents the first investigation
of cascading phenomena in networks with geometric constraints.

\section*{Appendix A}

The following lemma is similar to the one used in \cite{DoFrTh05,DoFrMaMeTh06,Gr99}. For
completeness, we provide the proof here.

\vspace{0.1in}%
\begin{lemma}\label{Lemma-Closed-Circuit-Number}
Given a square lattice $\mathcal{L}'$, suppose that the origin is located at the center of one
square. Let the number of circuits\footnote{A circuit in a lattice $\mathcal{L}'$ is a closed path
with no repeated vertices in $\mathcal{L}'$.} surrounding the origin with length $2m$ be
$\gamma(2m)$, where $m\geq 2$ is an integer, then we have
\begin{equation}\label{gamma-2m}
\gamma(2m)\leq \frac{4}{27}(m-1)3^{2m}.
\end{equation}
\end{lemma}
\vspace{0.1in}%
\begin{figure}[t!]
\centering
\includegraphics[width=2.5in]{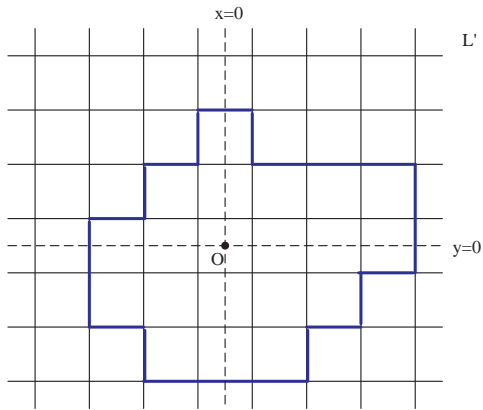}
\caption{An example of a circuit surrounding the origin in lattice
$\mathcal{L}'$}\label{fig:ClosedCircuit}
\end{figure}

\emph{Proof:} In Figure~\ref{fig:ClosedCircuit}, an example of a circuit that surrounds the origin
is illustrated. First note that the length of such a circuit must be even. This is because there
is a one-to-one correspondence between each pair of edges above and below the line $y=0$, and
similarly for each pair of edges at the left and right of the line $x=0$. Furthermore, the
rightmost edge can be chosen only from the lines $l_i: x=i-\frac{1}{2}, i=1,...,m-1$. Hence the
number of possibilities for this edge is at most $m-1$. Because this edge is the rightmost edge,
each of the two edges adjacent to it has two choices for its direction. For all the other edges,
each one has at most three choices for its direction. Therefore the number of total choices for
all the other edges is at most $3^{2m-3}$. Consequently, the number of circuits that surround the
origin and have length $2m$ must be less or equal to $(m-1)2^23^{2m-3}$, and hence we have
(\ref{gamma-2m}). \qed

\section*{Appendix B}

By~\eqref{p-1}, the probability $p_1$ can be written as
\begin{eqnarray}\label{eq:p-1-1}
p_1 & = & \sum_{k=1}^{\infty}\Pr(N(S_a)=k)\Pr(N'(S_a)\geq 1|N(S_a)=k) \nonumber\\
& = & \sum_{k=1}^{\infty}\Pr(N(S_a)=k)[1-\Pr(N'(S_a)=0|N(S_a)=k)] \nonumber\\
& = & \sum_{k=1}^{\infty}\frac{\left(\frac{\lambda}{2}\right)^k}{k!}e^{-\frac{\lambda}{2}}
[1-\Pr(I_1=1,...,I_k=1|N(S_a)=k)]
\end{eqnarray}
where $I_i$ is the indicator random variable indicating the failure of the $i$-th node. Because
$q(k)$ is non-decreasing in $k$, the event $\{I_i=1|N(S_a)=k\}$ is an increasing event. Hence,
according to the FKG inequality,
\begin{equation}\label{eq:FKG-1}
\Pr(I_1=1,...,I_k=1|N(S_a)=k)\geq [\Pr(I_i=1|N(S_a)=k)]^k.
\end{equation}
Since $d=\frac{\sqrt{2}}{2}$, all the nodes of $G(\mathcal{H}_\lambda,1)$ in $S_a$ are adjacent to
each other. Hence if there are $k$ nodes in $S_a$, every node of $G(\mathcal{H}_\lambda,1)$ in
$S_a$ has degree greater than or equal to $k-1$. In addition, since $q(k)$ is non-decreasing in
$k$, we have
\begin{equation}\label{eq:q-k-1}
\Pr(I_i=1|N(S_a)=k)\geq q(k-1).
\end{equation}
By~\eqref{eq:p-1-1}--\eqref{eq:q-k-1}, we have
\begin{eqnarray}\label{eq:p-1-1-result}
p_1 & \leq & \sum_{k=1}^{\infty}\frac{\left(\frac{\lambda}{2}\right)^k}{k!}e^{-\frac{\lambda}{2}}
\left(1-q(k-1)^k\right)
\nonumber\\
& = &
1-e^{-\frac{\lambda}{2}}-\sum_{k=1}^{\infty}\frac{\left(\frac{\lambda}{2}\right)^k}{k!}e^{-\frac{\lambda}{2}}
q(k-1)^k.
\end{eqnarray}

\section*{Appendix C}

\begin{figure}[t!] \centering
\includegraphics[width=1.5in]{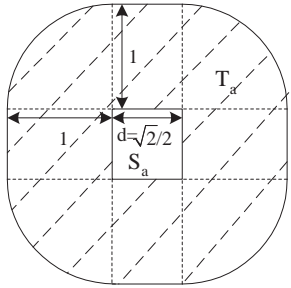}
\caption{$T_a$ and $S_a$}\label{fig:Ta}
\end{figure}

Let $T_a$ be the shaded area shown in Figure~\ref{fig:Ta}. Then $|T_a|=2\sqrt{2}+\pi$. Let
$N(T_a)$ be the number of nodes of $G(\mathcal{H}_\lambda,1)$ in $T_a$. Since $S_a$ and $T_a$ do
not overlap, $N(S_a)$ and $N(T_a)$ are independent.
By~\eqref{p-1}, we can write $p_1$ as
\begin{eqnarray}\label{eq:p-1-2}
p_1 & = & \sum_{k=1}^{\infty}\Pr(N(S_a)=k)\sum_{m=0}^{\infty}\Pr(N(T_a)=m)
\Pr(N'(S_a)\geq 1|N(S_a)=k, N(T_a)=m)\nonumber\\
& = & \sum_{k=1}^{\infty}\Pr(N(S_a)=k)\sum_{m=0}^{\infty}\Pr(N(T_a)=m)
[1-\Pr(N'(S_a)=0|N(S_a)=k, N(T_a)=m)]\nonumber\\
& = & \sum_{k=1}^{\infty}\frac{\left(\frac{\lambda}{2}\right)^k}{k!}e^{-\frac{\lambda}{2}}
\sum_{m=0}^{\infty}\Pr(N(T_a)=m) [1-\Pr(I_1=1,...,I_k=1|N(S_a)=k, N(T_a)=m)],
\end{eqnarray}
where $I_i$ is the indicator random variable indicating the failure of the $i$-th node. Because
$q(k)$ is non-increasing in $k$, the event $\{I_i=1|N(S_a)=k,N(T_a)=m\}$ is a decreasing event.
Hence, according to the FKG inequality,
\begin{equation}\label{eq:FKG-2}
\Pr(I_1=1,...,I_k=1|N(S_a)=k,N(T_a)=m)\geq [\Pr(I_i=1|N(S_a)=k, N(T_a)=m)]^k.
\end{equation}
For any node $u$ inside $S_a$, all of $u$'s neighbors are within $T_a\cup S_a$. Given $N(S_a)=k$
and $N(T_a)=m$, any node $u$ inside $S_a$ has degree less than or equal to $m+k-1$. In addition,
since $q(k)$ is non-increasing in $k$, we have
\begin{equation}\label{eq:q-k-2}
\Pr(I_i=1|N(S_a)=k, N(T_a)=m)\geq q(m+k-1).
\end{equation}
By~\eqref{eq:p-1-2}--\eqref{eq:q-k-2}, we have
\begin{eqnarray}\label{p-1-result-2}
p_1 &\leq& \sum_{k=1}^{\infty}\frac{\left(\frac{\lambda}{2}\right)^k}{k!}e^{-\frac{\lambda}{2}}
\sum_{m=0}^{\infty}\Pr(N(T_a)=m) \left(1-q(m+k-1)^k\right)\nonumber\\
&=& \sum_{k=1}^{\infty}\frac{\left(\frac{\lambda}{2}\right)^k}{k!}e^{-\frac{\lambda}{2}}
\sum_{m=0}^{\infty}\frac{[\lambda(2\sqrt{2}+\pi)]^m}{m!}e^{-\lambda(2\sqrt{2}+\pi)}
\left(1-q(m+k-1)^k\right).
\end{eqnarray}

\bibliography{PercolationTopic2}
\bibliographystyle{ieeetr}

\end{document}